\documentclass[printer]{aa}  

\usepackage{graphicx}
\usepackage[scientific-notation=true]{siunitx}
\usepackage{multirow}
\usepackage{ulem}
\usepackage{subcaption}
\usepackage{xcolor}
\usepackage[usestackEOL]{stackengine}
\edef\tmp{\the\baselineskip}
\usepackage{amsmath}
\usepackage{stfloats}
\usepackage{txfonts}

\usepackage[breaklinks,colorlinks,citecolor=blue]{hyperref}

\begin{document}

\title{Accretion variability in RU\,Lup
    \thanks{Based on observations obtained at the Canada-France-Hawaii Telescope (CFHT) and the Cerro Tololo Inter-American Observatory (CTIO) Small and Medium Aperture Research Telescope Facility (SMARTS).}
    \thanks{The flux calibrated ESPaDOnS data shown in Fig.~\ref{fig:full_spec} and Tables \ref{table:laccandmacc}, \ref{table:H_fluxes}, and \ref{table:HeI_fluxes} are available in electronic form at the CDS via anonymous ftp to \url{cdsarc.u-strasbg.fr} (130.79.128.5) or via\\\url{http://cdsweb.u-strasbg.fr/cgi-bin/qcat?J/A+A/}.}}
    

   \author{C. Stock
          \inst{1}\fnmsep\inst{2},
          P. McGinnis\inst{1},
          A. Caratti o Garatti\inst{1}\fnmsep\inst{3},
          A. Natta\inst{1}
          \and
          T. P. Ray\inst{1}\fnmsep\inst{2}
          }

   \institute{Dublin Institute for Advanced Studies (DIAS), School of Cosmic Physics, Astronomy and Astrophysics Section, 31 Fitzwilliam Place, Dublin 2, Ireland\\
              \email{cstock@cp.dias.ie}
         \and Trinity College Dublin, School of Physics, College Green, Dublin 2, Ireland
         \and INAF - Osservatorio Astronomico di Capodimonte, Salita Moiariello 16, 80131 Napoli, Italy
             }

   \date{Accepted 23 Sept. 2022}
   
    \abstract
    {The process of accretion in classical T Tauri stars (CTTSs) has been observed to vary on different timescales. Studying this variability is vital to understanding a star's evolution and provides insight into the complex processes at work within, including sources of the veiling present. Understanding the dichotomy between continuum veiling and emission line veiling is integral to accurately measuring the amount of veiling present in stellar spectra.} 
    {Here, 15 roughly consecutive nights of optical spectroscopic data from the spectropolarimeter ESPaDOnS are utilised to characterise the short-term accretion activity in the CTTS, \object{RU\,Lup}, and investigate its relationship with the veiling in the \ion{Li}{i}\,6707\,\AA\,absorption line.}
    {The accretion-tracing \ion{H}{i} Balmer series emission lines were studied and used to obtain the accretion luminosity ($L_{\text{\,acc}}$) and mass accretion rate ($\dot{M}_{\text{\,acc}}$) for each night, which vary by a factor of $\sim\!2$ between the brightest and dimmest nights. We also measured the veiling using multiple photospheric absorption lines (\ion{Na}{i}\,5688\,\AA, \ion{Mn}{i}\,6021\,\AA, and \ion{Li}{i}\,6707\,\AA) for each night.} 
    {We find the \ion{Li}{i}\,6707\,\AA\,line provides measurements of veiling that produce a strong, positive correlation with $L_{\text{\,acc}}$ in the star. When corrected for Li depletion, the average veiling measured in the \ion{Li}{i}\,6707\,\AA\,line is $r_{\,\ion{Li}{i},\,\text{avg}}\sim\!3.25\pm0.20$, which is consistent with the other photospheric lines studied ($r_{\text{avg}}$\,$\sim\!3.28\pm0.65$).}
    {We measured short timescale variability in the accretion luminosity and mass accretion rate that are intrinsic and not due to geometric effects. As the forbidden line emission we observe ([\ion{O}{i}]\,6300\,\AA\,and [\ion{S}{ii}]\,6730\,\AA) remains remarkably constant over our epochs, it is clear that the variations in the mass accretion rate are too short to have an effect on these outflow tracers. Upon comparing the changes in veiling and accretion luminosity, we find a strong, positive correlation. This study provides an example of how this correlation can be used as a tool to determine whether a measured variability is due to extinction or an intrinsic change in accretion. As the determination of veiling is an independent process from measuring $L_{\text{\,acc}}$, their relationship allows further exploration of accretion phenomena in young stars.}

               
   \titlerunning{Accretion variability in RU\,Lup}
   \authorrunning{Stock, C. et al.}
   
\maketitle
\section{Introduction}
\label{intro}

Many processes at work in low-mass ($M_*\leq2$\,M$_\sun$) young stars influence their formation and evolution. Two of the most important ones are accretion and ejection of material onto and emanating from young stars, such as T Tauri stars (TTSs). These TTS are low-mass pre-main sequence stars which are commonly divided into two main classes: classical T Tauri stars (CTTSs), which are younger ($\sim$10$^6$\,yr) and more actively accreting than weak-line T Tauri stars (WTTS), which are often older ($\sim$10$^7$\,yr) and at the end of the accretion phase \citep[][and references therein]{Walter1988,Bertout1989,Hartmann2016}. The CTTSs demonstrate a rich emission spectrum mostly due to accretion of material theorised to be structured as funnels, flowing along the stellar magnetic field lines from the inner disk to the stellar surface \citep[e.g.][and references therein]{Hartmann2016}. Other spectral lines come from the ejection phenomena. They are varied in their structures (e.g. jets and winds) and velocities, depending on the conditions from which they originate; however, it has been shown that a strong link between ejection and accretion exists~\citep[e.g.][and references therein]{Ray07}.

The accretion luminosity ($L_{\text{\,acc}}$) of a young star is produced by the shocked material falling onto the stellar photosphere. The accretion energy released is seen in the form of continuum excess in emission  -- mostly at ultraviolet (UV) and optical wavelengths -- and emission lines. Historically, the $L_{\text{\,acc}}$ of low-mass young stars is estimated by modelling the Paschen and Balmer continua emission as isothermal slabs \citep{Valenti1993, Gullbring1998}. More often, $L_{\text{\,acc}}$ has been computed from empirical relations
between the luminosity of accretion-tracing lines and $L_{\text{\,acc}}$~\citep[see e.g.][]{Gulbring1998,Muzerolle1998,Natta2006}.
\citet{Alcala2017} derived more precise relationships between the line luminosity ($L_{\,\text{line}}$) of several optical and near-infrared (NIR) accretion-tracing emission lines and $L_{\text{\,acc}}$. 
 
This physical relationship between $L_{\,\text{line}}$ and $L_{\text{\,acc}}$ allows the estimation of the mass accretion rate ($\dot{M}_{\text{\text{\,acc}}}$) \citep{Hartmann1998}. The $\dot{M}_{\text{\,acc}}$ of CTTSs was calculated by \citet{Herczeg2008} by analysing the excess UV and optical emission. A decade later, estimates of the $\dot{M}_{\text{\,acc}}$ for a large survey of young stars by \citet{Alcala2017} was derived. Accretion through the disk is studied widely through shock models as well \citep[e.g.][]{Dodin2012,Gullbring1998,Espaillat2021}.

Accretion in young stars is not a steady process, but rather variable~\citep[for a review see][and references therein]{Fischer2022}. Understanding how young stars' accretion activity varies is vital for determining how they gain their mass and what implications that process has on their evolution \citep{Herbst1994}.

\begin{table*}[ht]
\centering
\begin{tabular}{l c c r}
\hline\hline  
\noalign{\smallskip}
Distance & $158.9 \pm 0.7$ & pc & \citet{GaiaCollaboration2018} \\ 
Age & $2-3$ & Myr & \citet{Herczeg2005} \\
$M_{\star}$ & $\sim\!0.65$ & $M_{\odot}$ & \citet{Herczeg2005} \\
$R_{\star}$ & $\sim\!1.64$ & $R_{\odot}$ & \citet{Herczeg2005} \\
SpT & K7 & - & \citet{Bailer-Jones2018} \\
$T_{\text{eff}}$ & $4037\pm96$ & K & \citet{Frasca2017} \\
$L_{\star}$ & $1.313 \pm 0.605$ & $L_{\odot}$ & \citet{Alcala2017} \\
$P_{\text{rot}}^*$ & $3.71 \pm 0.01$ & days & \citet{Stempels2007} \\
$v$ sin\,$i$ & $8.5 \pm 4.8$ & km s$^{-1}$ & \citet{Frasca2017} \\
$A_V$ & $\sim\!0.07$ & mag & \citet{Herczeg2005} \\
$i^{**}$ & $16\,\pm\,^{6}_{8}$ & $\degr$ & \citet{GRAVITY}\\
\noalign{\smallskip}
\hline
\noalign{\smallskip}
\end{tabular} \\
\footnotesize{$*$ Calculated from radial velocity measurements; $^{**}$ Disk inclination with respect to the plane of the sky}
\caption{Stellar properties of RU\,Lup.}
\label{table:stellar_props}
\end{table*}

Indeed, luminosity amplitude and timescales vary depending on the origin of such variability~\citep[see Figure~3 of][]{Fischer2022}. Temporal changes over timescales of days and amplitudes less than one order of magnitude in the accretion rate (the so-called routine variability) are often attributed to the rotation of the star and are very useful in measuring its rotation period \citep[e.g.][]{Scholz2006,Costigan2012,Costigan2014,Rebull2014}.
However, this is not exclusively the case as the rotation contribution might hide any intrinsic changes that come from the stellar-disk magnetospheric interaction.
On the other hand, the so-called accretion bursts with larger amplitude (more than one order of magnitude) and timescale variability (from months to years) are more related to the disk's mass transport and refurbishment, and they might originate from disk instabilities and other related mechanisms relevant to EXor and FUor phases \citep{Hartmann2016}.

Therefore, studying the variation of $L_{\text{\,acc}}$ and $\dot{M}_{\text{\,acc}}$ is important in understanding how TTSs form. Magnetospheric accretion can be divided into two regimes: stable and unstable. These two regimes were predicted by several 3D magnetohydrodynamic (MHD) simulations of disk accretion onto a rotating young star \citep{Romanova2008, Kulkarni2008, Kulkarni2009}. According to these models, the main factors that determine in which regime a star accretes are the mass accretion rate, the strength of the magnetic field, the stellar rotation rate, and the misalignment between the star’s rotation axis and its magnetic poles. 

Classical T Tauri star accretion activity and variability can also be studied by using veiling. This is the filling-in of stellar photospheric absorption features such that they appear shallower in observed spectra of an actively accreting young star, due to an accretion-fueled flux excess (both in the continuum and in lines). This phenomenon has been investigated since \citet{Joy1949} first introduced the concept of veiling. With high resolution spectroscopy, it is possible to observe the photospheric absorption lines of the star and this excess due to accretion that effectively veils the spectrum.

This study focusses on the CTTS \object{RU\,Lup}, which is known to have a plethora of strong accretion features \citep[][and references therein]{Herczeg2005, Gahm2008, Siwak2016}. RU\,Lup is an accreting CTTS of 2 -- 3 Myrs that lies in the Lupus star-forming region at a distance of $d\approx159$ pc \citep{GaiaCollaboration2018}. It has a rotational period of $3.71\pm0.01$ \citep{Stempels2007} and is notable for its strong emission lines and veiling. These accretion-tracing lines are exceptionally broad, for example, the equivalent width (EW) of the H$\alpha$ emission line is usually found to be greater than 100\,\AA\,\citep{Siwak2016}. In their models, this high accretion activity was found by \citet{Lamzin1996} to be responsible for the majority of the observed luminosity, whereby only about 10\% is generated by gravitational contraction of the star itself. The $\dot{M}_{\text{\,acc}}$ was calculated by \citet{Herczeg2008} and found to be $\sim\!1.8\times10^{-8}$ $M_{\odot}$yr$^{-1}$. The study by \citet{Alcala2017} resulted in values of $\sim\!6.7\times 10^{-8}$ $M_{\odot}$yr$^{-1}$. This spread is indicative of the amount of accretion variability exhibited in RU\,Lup on the timescales of years. For this study we measured the $\dot{M}_{\text{\,acc}}$ of RU\,Lup during 2011 and determined how it changes over the shorter timescale of two weeks. By investigating its short-time stellar-disk magnetospheric interaction, our dataset allowed the exploration of how RU\,Lup's accretion fits into the category of routine variability, as delineated in Table 1. of \citet{Fischer2022}. 

Additionally, multiple studies have inferred that RU\,Lup has a near face-on geometry from varying methods including the analysis of line profiles and measured rotational velocities \citep[see \S\,2 of][and references therein]{Herczeg2005} or the direct measurement of the inner dusty disk geometry and size with NIR interferometry~\citep{GRAVITY}. Because it is seen mostly face-on, RU\,Lup is an excellent candidate for a study of accretion variability as there are fewer geometrical effects to confound the observations (such as stellar rotation and extinction from the disk). It is also a well-studied and interesting object undergoing strong accretion activity. Stellar properties of RU\,Lup such as its radius ($R_{\star}$), spectral type (SpT), stellar luminosity ($L_{\star}$), rotational period ($P_{rot}$), $v$\,sin\,$i$, and visual extinction ($A_V$) can be found in Table \ref{table:stellar_props}.

\section{Observations and data reduction}
\label{section:observations}

\subsection{Spectroscopy}
High-resolution optical spectroscopic data were obtained with the 3.6\,m Canada-France-Hawaii Telescope (CFHT) Echelle SpectroPolarimetric Device for the Observation of Stars (ESPaDOnS) across 12 epochs from 08 June 2011 to 23 June 2011 (proposal ID: 11AP12, PI: Jean-Francois Donati). CFHT/ESPaDOns is a fibre-fed echelle spectrograph and spectropolarimeter that operates across the entire optical wavelength range (370\,nm to 1050\,nm). These observations were taken in `object + sky' mode with a spectral resolution of $\sim$68,000 and spectral sampling bin size of 1.8 km\,s$^{-1}$. 

The ESPaDOnS data were reduced with `Upena', the standard pipeline provided by the CFHT, based on the Libre-ESpRIT package \citep{Donati1997}. Additionally, because no standard star observation was available, the telluric lines in the regions of interest to this study were fitted and removed manually.

This included the region in and around the [\ion{O}{i}]\,6300\,\AA\,line, that is the stronger component of the doublet that also includes [\ion{O}{i}]\,6364\,\AA, containing strong telluric contamination. This was done by first fitting and subtracting the continuum from the spectra in this region, and then, utilising libraries such as Astropy in Python, each telluric absorption that was blended with the [\ion{O}{i}]\,6300\,\AA\,line emission was identified and fitted with a Gaussian. These Gaussian fits were then subtracted from each night's spectrum. 

\begin{figure*}[ht]
    \centering
    \includegraphics[width=1.0\linewidth]{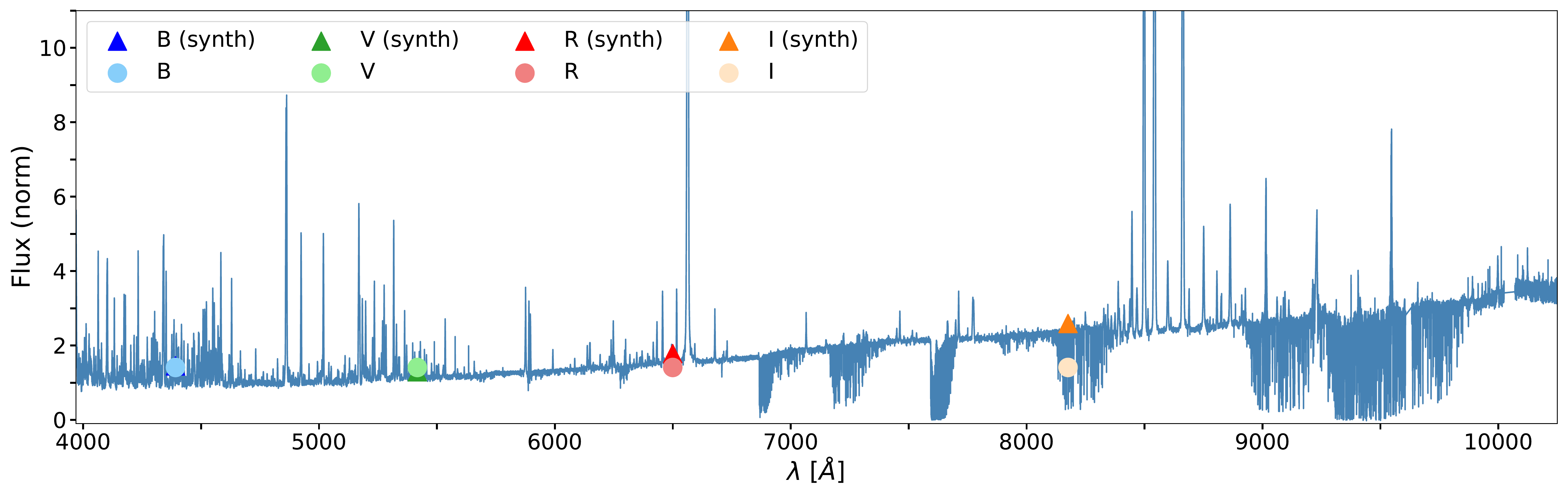}
    \caption{ESPaDOnS optical spectrum of RU\,Lup on the night of 17 June 2011. This spectrum is overlaid with the photometry measured from the ANDICAM observations of the same night. The circles represent the aperture photometry taken in the $B$, $V$, $R$ and $I$ bands and the triangles represent the synthetic photometry calculated by convolving the spectrum with the transmission response of ANDICAM in each band. For the $B$ and $V$ bands the triangles and circles are overlapping.} 
    \label{fig:sed_and_photo}
\end{figure*}

In addition to telluric contamination, the data contain many photospheric absorption lines, typical of a K7 star. Some of these were blended with the [\ion{O}{i}]\,6300\,\AA\,line and required removal to study its profile. Unlike the telluric absorption lines, it was possible to remove all the photospheric lines in this region ($6284 - 6342$\,\AA) at once. This was done by obtaining a WTTS spectrum of the same spectral type as a template, which contains the same photospheric lines but no emission lines \citep{Martin1998}, and subtracting this from the CTTS spectrum of RU\,Lup. As WTTSs undergo little-to-no accretion activity (which produces many emission lines), their spectra can be used as templates for the photosphere of CTTSs of the same spectral type.
Indeed, the RU\,Lup spectra was compared with the WTTS spectrum of TAP\,45 \citep[same spectral type and similar $v$\,sin\,$i$\,=\,$7.7\pm0.7$\,km\,s$^{-1}$;][]{Nguyen2012}. TAP\,45 has a stellar mass of $\sim\!0.79\,M_{\odot}$, a radius of $\sim\!1.14\,R_{\odot}$ and $T_{\text{eff}}\sim\,4040$\,K \citep[][]{Grankin2013}. This template was then re-scaled such that it met the level of the RU\,Lup continuum; these parameters were adjusted by minimising a $\chi^2$ fit. The resultant broadened and veiled template represents a spectrum containing the fitted photospheric features which can then be used to subtract the contaminating photospheric lines from the regions of interest. This process was completed on a night-by-night basis.

Finally, the radial velocities of the observed lines were calculated using single or multiple Gaussian fits (the detected lines typically show multiple components). These velocities are measured with respect to the systematic velocity of the star \citep[$v$~=~$+8~\pm~2$~km~s$^{-1}$,][]{Takami2001} in the local standard of rest (LSR) frame.

\subsection{Photometry and spectroscopic flux calibration}
\label{subsection:flux_calib}

As the ESPaDOnS observations were not flux calibrated, we searched archival data and found simultaneous photometry of RU Lup from the 1.3\,m SMARTS/CTIO\footnote{SMARTS, the Small and Medium Aperture Research Telescope Facility, is a consortium of universities and research institutions that operate the small telescopes at Cerro Tololo under contract with AURA.} telescope for the night of 17 June 2011. These ANDICAM photometric data were taken on 17 June 2011, allowing the ESPaDOnS spectra from the same night to be flux calibrated. The data from the $B, V, R$ and $I$ filters were reduced in Python using the `Astropy' and `ccdproc' packages to apply biases, darks and flat-fielding (dome flats for $V, R$ and $I$ filters and sky flats for the $B$ filter) to the raw images of RU Lup and the standard stars observed from the same night. The resulting standard stars were flux calibrated using the zero-point fluxes given for the Kitt Peak National Observatory (KPNO) filters used on ANDICAM, and were then used to flux calibrate the science images. The calibrated spectrum of the 17 June 2011 night is shown in Fig.\,\ref{fig:sed_and_photo}, together with the photometric data. 

The ESPaDOnS instrument provides spectroscopic data from the $B$ to the $I$ band ($3700-10500$\,\AA). We have computed the synthetic photometry for these bands, using the transmission response of ANDICAM (triangles in Fig. \ref{fig:sed_and_photo}). While the $B$ to $R$ synthetic photometry matches the aperture photometry (circles) well, the $I$ band synthetic photometry is about a factor of two higher than the measured value. We believe this to be due to an instrumental issue, and therefore have limited the spectral range we use in this study to the $B$ through $R$ region, where the data are well-calibrated.

\begin{figure*}[b]
    \centering
    \includegraphics[width=0.8\linewidth]{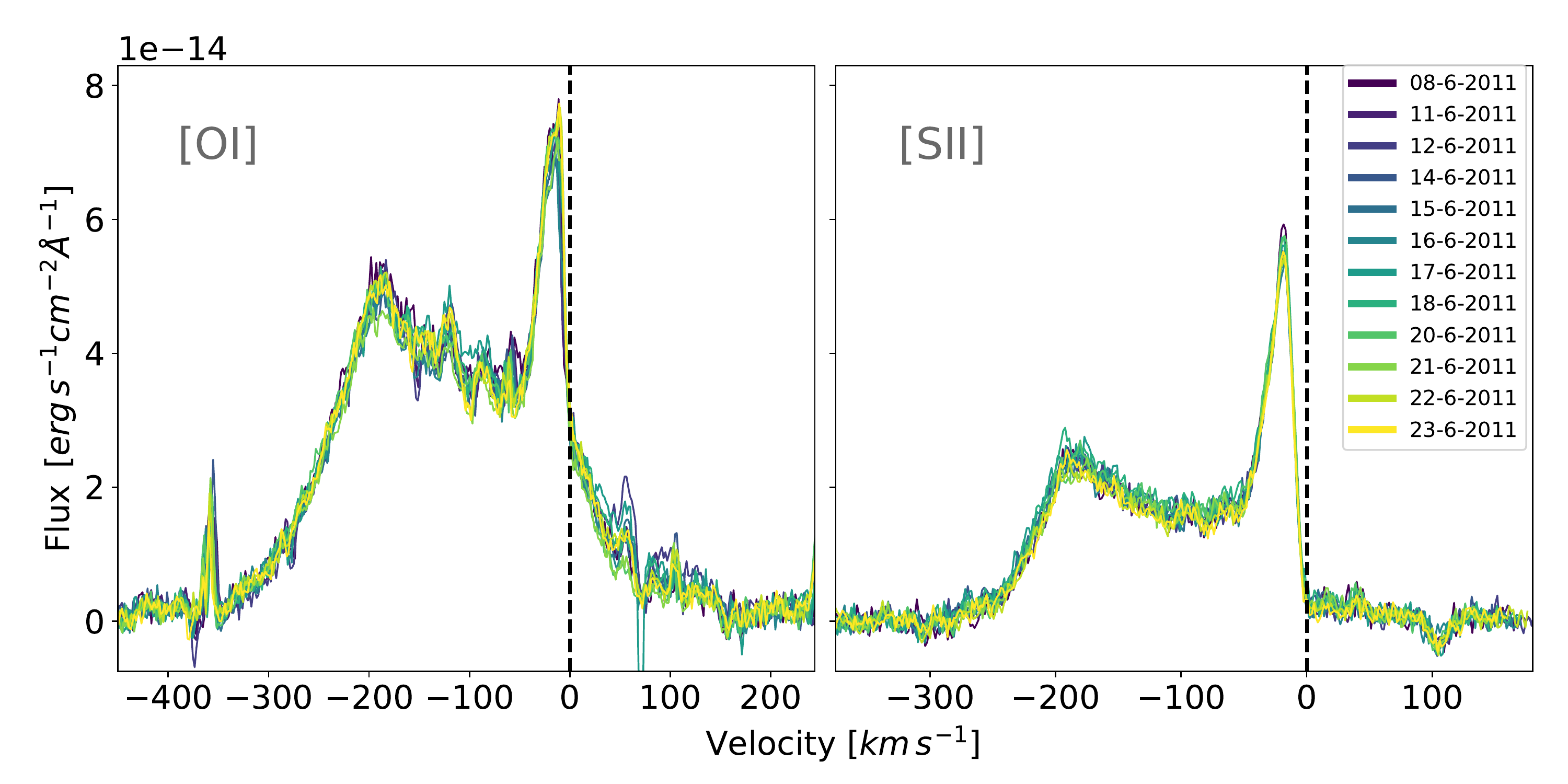}
    \caption{Outflow-tracing lines of [\ion{O}{i}]\,6300\,\AA\,and [\ion{S}{ii}]\,6730\,\AA\,over the 12 epochs of ESPaDOnS observations spanning from 08 June 2011 to 23 June 2011. These spectra are continuum-subtracted and flux calibrated. The features at $\sim\!-370$\,km\,s$^{-1}$ and $\sim+80$\,km\,s$^{-1}$ are residuals from the telluric removal.}
    \label{fig:outflow_lines}
\end{figure*}

To the best of our knowledge, no photometric data are available for the other nights. Therefore, we looked to the [\ion{S}{ii}]\,6730\,\AA\, emission line which has extremely similar profiles from night-to-night, and very small night-to-night flux variations. The profiles of this line show the typical features of forbidden lines in CTTS, namely a high- and a low-velocity component (HVC and LVC, respectively) wherein the HVC is not expected to show any variability on these short timescales (see \S\,\ref{subsection:forbidden_lines}) \citep{Hartigan1995, Rigliaco2013, Simon2016, McGinnis2018}. Taking the flux calibrated [\ion{S}{ii}]\,6730\,\AA\, HVC (the 17 June epoch) as a reference (the [\ion{O}{i}]\,6300\,\AA\, line can also be used, but is less straightforward to analyse because it is in a region of telluric contamination), we re-scaled each epoch by the amount that was necessary to match the flux of its HVC with the flux calibrated one. 

The lack of temporal variability seen in the resultant [\ion{O}{i}]\,6300\,\AA\, line spectra (Fig.\,\ref{fig:outflow_lines}, left panel) confirms that our re-scaling is a valid calibration, as the [\ion{O}{i}]\,6300\,\AA\, line's HVC is not expected to vary over these timescales either. We estimate that the uncertainty of the calibrated flux is of the order of 5\%.

In the specific case of RU\,Lup, our choice is also supported by the results of \citet{Gahm2013}, which found that the forbidden line profiles remained remarkably constant over their study's longer timescale of five years. The authors also note that in their flux  calibrated spectra, the flux of the [\ion{S}{ii}]\,6730\,\AA\, line remains stable for relatively moderate veiling \citep[$r\leq 2$; see Figure~2 in][]{Gahm2008}. Indeed, this is the case in our data set (see \S\,\ref{subsection:veiling_metals}).

Our assumptions separately work on both [\ion{O}{i}]\,6300\,\AA\, and [\ion{S}{ii}]\,6730\,\AA\,line flux calibration. Our calibrated spectra are presented in Fig.\,\ref{fig:full_spec}, where the main lines used in this paper are labelled. The [\ion{S}{ii}]\,6730\,\AA\,emission line is the stronger component of the \ion{S}{ii} doublet that also includes [\ion{S}{ii}]\,6716\,\AA.

\section{Results}
\label{results}

\begin{figure*}
    \centering
    \includegraphics[width=1.0\linewidth]{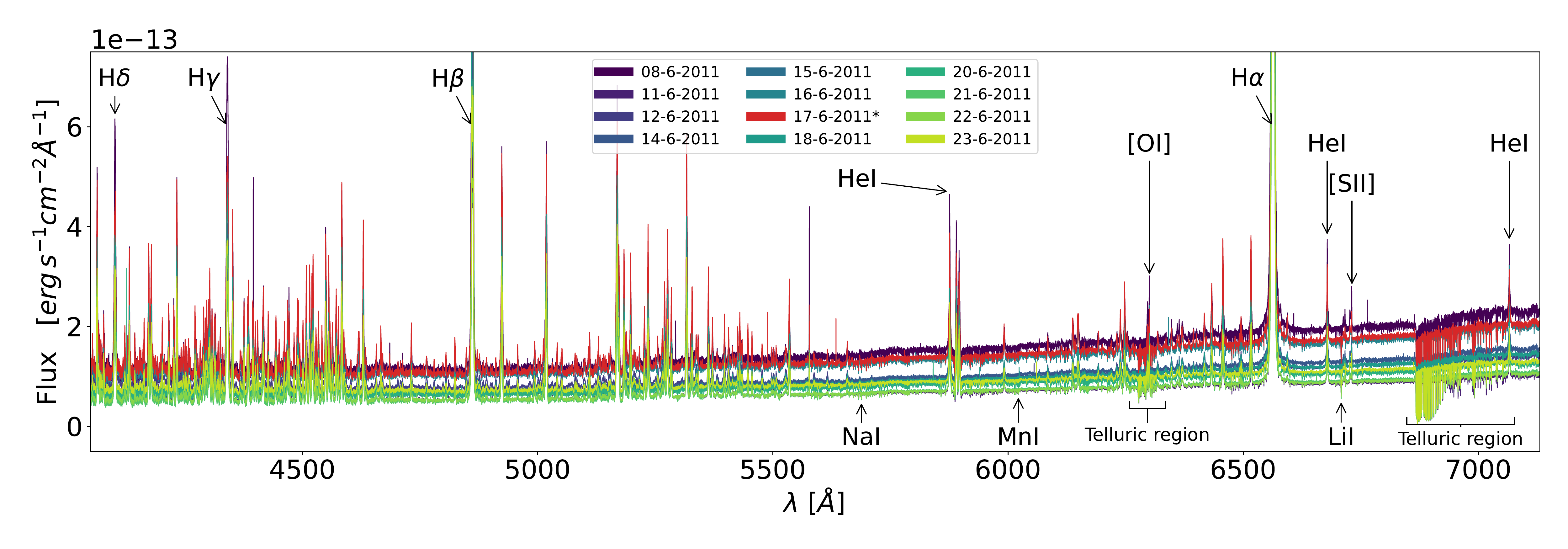}
    \footnotesize{* simultaneous with ANDICAM photometry}
    \caption{Flux calibrated spectra from the 12 epochs taken with ESPaDOnS across the optical regime ($4000 - 7100$\,\AA). Notable emission and absorption lines discussed in this study are identified with their species name. Regions highly contaminated by telluric emission are also indicated. The spectrum in black (17-06-2011) represents the night of observations that we calibrated with simultaneous photometry. The other 11 nights were flux calibrated as described in \S\,\ref{subsection:flux_calib}.}
    \label{fig:full_spec}
\end{figure*}

Our multi-epoch calibrated spectra are presented in Fig.\,\ref{fig:full_spec}. The spectra, from the $B$ to the $R$ band, show variations in the continuum flux. Many photospheric lines in absorption are present, indicative of a K7 star \citep{White2003}. Among them, the \ion{Li}{i}\,6707\,\AA\,transition is the most prominent. Additionally, we detect an overwhelming number of emission lines, typically originating from the circumstellar environment of RU\,Lup such as accretion shocks, winds, jet, and chromosphere. For the purposes of this paper, we select a number of lines that are widely used to trace accretion and ejection activity in CTTS.

\subsection{Accretion-tracing lines}
\label{subsection:acc_lines} 

As previously mentioned, the accretion activity in young stars such as RU\,Lup produces line emission that allows the study of these processes. There are many known accretion-tracing emission lines at optical wavelengths, and their intensities, profiles, and relative red- and blue-shifts provide information about their origins. 

The brightest accretion-tracing lines present in our data are \ion{H}{i} (H$\alpha$, H$\beta$, H$\gamma$ and H$\delta$) and \ion{He}{i} (5875\,\AA, 6678\,\AA, and 7065\,\AA). Equivalent widths and integrated fluxes are shown in Tables \ref{table:H_fluxes} and \ref{table:HeI_fluxes}, respectively. Both line profiles and intensities vary with species and (most noticeably in the case of \ion{H}{i}) from night to night (see Figure~\ref{fig:acc_lines}). In the following we concentrate on the variation of the line luminosity.

As seen in the left-hand column of Fig.\,\ref{fig:acc_lines}, the Balmer lines vary significantly over the short timescale of two weeks covered by the ESPaDOnS data. Between the dimmest and brightest nights, there is an increase in integrated flux by a factor of $\sim$1.7 for H$\alpha$, $\sim$1.9 for H$\beta$, $\sim$2.2 for H$\gamma$ and $\sim$2.0 for H$\delta$. 

The right-hand column of Fig.\,\ref{fig:acc_lines} shows the \ion{He}{i} emission lines at 5875\,\AA, 6678\,\AA\,and 7065\,\AA. The trend in flux variability seen here is similar to that for the Balmer series. There is an increase in the integrated intensity between the dimmest and brightest nights by a factor of $\sim\,$1.9 for \ion{He}{i}\,5875\,\,\AA\,and $\sim\,2$ for \ion{He}{i}\,6678\,\AA\,and \ion{He}{i}\,7065\,\AA. 

As has been shown by \citet{Beristain2001}, the \ion{He}{i}\,5875\,\AA\,line can be decomposed into a narrow component (NC) and a broad component (BC). The NC is generally slightly red-shifted and is believed to originate in the post-shock region near the stellar surface. The BC has been theorised to have multiple origins, with contributions from the accretion columns as well as from a hot wind. We have decomposed the \ion{He}{i}\,5875\,\AA\,and \ion{He}{i}\,6678\,\AA\,lines into two components in our data. Though the components vary from night to night, both profiles indeed show a clear NC and BC every night (Fig. \ref{fig:heI_decomp}). The flux of each component varies with a similar trend, indicating that the processes that give rise to both components correlate well with each other, and therefore with accretion.

\subsection{Time-dependent accretion luminosity \& mass accretion rate}
\label{subsection:macc_var}  

We derive $L_{\text{\,acc}}$ from the luminosity of the accretion-tracing lines, using the well established relations between each line luminosity ($L_{\,\text{line}}$) and $L_{\text{\,acc}}$ derived in \citet{Alcala2017} for CTTSs. $L_{\text{\,acc}}$ is then computed for each night from the average of the $L_{\text{\,acc}}$ values derived from the individual lines (see Fig.\,\ref{fig:man_plot}). We note that because of the way the $L_{\,\text{line}}\,vs.\,L_{\text{\,acc}}$ relation was derived in \citet{Alcala2017}, $L_{\text{\,acc}}$ refers to the continuum excess emission only.

The inferred $L_{\text{\,acc}}$ values found in the \ion{He}{i} differ slightly from those of the \ion{H}{i} lines. This is likely due to the complexity of processes that the \ion{He}{i} lines are tracing, which include accretion, but not exclusively. In particular there can be a strong wind emission \citep{Beristain2001,Johns-Krull2001,McGinnis2020}. Therefore we have decided to estimate $L_{\text{\,acc}}$ using the \ion{H}{i} emission lines only. 

Our values for $L_{\text{\,acc}}$ range from 0.35 to 0.72 \,L$_{\odot}$ over the course of 15 days, with an average of $0.5\,L_{\odot}$ (see Table \ref{table:laccandmacc}). It is worth noting that the observed variability in the accretion luminosity is real, as the line flux variation of the different \ion{H}{i} lines is much larger than their flux uncertainties (see Tables~\ref{table:H_fluxes} and \ref{table:HeI_fluxes}). 

Variations in $L_{\text{\,acc}}$ can in principle be due to variations in extinction. However, there is no evidence in the literature of significant variations in visual extinction which is always very small \citep[e.g.][]{Herczeg2005,Alcala2017} and consistent with this system being oriented face-on. Moreover, Fig. \ref{fig:man_plot} shows there is no trend of increasing $L_{\text{\,acc}}$ from lines of increasing wavelength. Therefore, we assume in the following that changes in $L_{\text{\,acc}}$ trace physical changes in the accretion rate.

Because this variability cannot be caused by changes in extinction, it must come from accretion variability. The derived range of $L_{\text{\,acc}}$ values is in line with those found in the literature, derived with different methods (e.g. slab modelling of the Balmer jump, luminosity of \ion{H}{i} lines, UV excess), namely $\sim$0.5\,$L_{\odot}$ \citep{Herczeg2005,Antoniucci2014,Alcala2017}. 

\begin{table*}
\centering
\begin{tabular}{l c c c c c c c c}
\hline\hline  
\noalign{\smallskip}
Date & log $L_{\text{\,acc}}$ & $\pm\,\sigma$ & log $\dot{M}_{\text{\,acc}}$ & $r\,_{\ion{Na}{i}}$ & $r\,_{\ion{Mn}{i}}$ & $r\,_{\ion{Li}{i}}$ (obs) & $r\,_{\ion{Li}{i}}$* & $\chi^{\,2}$ \\ 
\noalign{\smallskip}
yyyy-mm-dd & [$L_{\odot}$] & [dex] & [$M_{\odot}\,$yr$^{-1}$] &  &  &  &  & \\ 
\noalign{\smallskip}\hline
\noalign{\smallskip}
2011-06-08 & $-0.35$ & $0.13$ & $-7.36$ & $2.66 \pm 0.44$ & $2.23 \pm 0.46$ & 1.11 & $2.91 \pm 0.24$ & 8.7 \\
2011-06-11 & $-0.27$ & $0.13$ & $-7.28$ & $2.98 \pm 0.55$ & $3.56 \pm 0.35$ & 1.56 & $3.36 \pm 0.20$ & 4.8 \\
2011-06-12 & $-0.14$ & $0.11$ & $-7.16$ & $4.61 \pm 1.06$ &       **        & 2.51 & $4.31 \pm 0.19$ & 2.5 \\
2011-06-14 & $-0.36$ & $0.13$ & $-7.37$ & $2.21 \pm 0.32$ & $2.08 \pm 0.42$ & 1.04 & $2.84 \pm 0.23$ & 7.6 \\
2011-06-15 & $-0.32$ & $0.13$ & $-7.34$ & $3.53 \pm 0.72$ & $4.55 \pm 1.39$ & 1.48 & $3.28 \pm 0.16$ & 5.2 \\
2011-06-16 & $-0.36$ & $0.13$ & $-7.38$ & $3.94 \pm 0.86$ & $3.07 \pm 0.73$ & 1.50 & $3.30 \pm 0.17$ & 4.3 \\
2011-06-17 & $-0.31$ & $0.13$ & $-7.33$ & $3.90 \pm 0.79$ & $4.96 \pm 1.63$ & 1.79 & $3.59 \pm 0.22$ & 5.9 \\
2011-06-18 & $-0.21$ & $0.14$ & $-7.23$ & $3.20 \pm 0.58$ & $2.81 \pm 0.61$ & 1.37 & $3.17 \pm 0.20$ & 5.8 \\
2011-06-20 & $-0.38$ & $0.15$ & $-7.40$ & $2.86 \pm 0.48$ & $2.43 \pm 0.51$ & 1.06 & $2.86 \pm 0.18$ & 6.4 \\
2011-06-21 & $-0.46$ & $0.14$ & $-7.47$ & $2.45 \pm 0.39$ & $2.07 \pm 0.18$ & 0.93 & $2.73 \pm 0.20$ & 8.5 \\
2011-06-22 & $-0.28$ & $0.13$ & $-7.30$ & $3.11 \pm 0.54$ & $3.10 \pm 0.75$ & 1.33 & $3.13 \pm 0.15$ & 5.0 \\
2011-06-23 & $-0.28$ & $0.13$ & $-7.29$ & $3.84 \pm 0.81$ & $5.35 \pm 0.30$ & 1.71 & $3.51 \pm 0.24$ & 4.3 \\
\noalign{\smallskip}
\hline
\end{tabular} \\
\smallskip
\footnotesize{* $r\,_{\ion{Li}{i}}$ = $r\,_{\ion{Li}{i}}$ (obs) + $\Delta r\,_{\text{abund.\,corr.}}$, where $\Delta\,r\,_{\text{abund.\,corr.}} = 1.8$, see \S\,\ref{section:LiVeiling}}\\
\footnotesize{** signal-to-noise ratio (S/N) insufficient to measure veiling}
\caption{$L_{\text{\,acc}}$, $\dot{M}_{\text{\,acc}}$, and veiling for each epoch. The $L_{\text{\,acc}}$ and $\dot{M}_{\text{\,acc}}$ for each night are determined by averaging the values found from the \ion{H}{i} lines, and their uncertainties are estimated from their spread. Columns 5 and 6 contain the veiling measurements for the photospheric lines, \ion{Na}{i}\,5688\,\AA\,and \ion{Mn}{i}\,6021\,\AA. The observed veiling measured in the \ion{Li}{i}\,6707\,\AA\,photospheric line is given in column 7 ($r\,_{\ion{Li}{i}}$\,(obs)) and the abundance corrected veiling is given in column 8 ($r\,_{\ion{Li}{i}}$ = $r\,_{\ion{Li}{i}}$ (obs) + $\Delta r\,_{\text{abund.\,corr.}}$). The final column contains the $\chi^{\,2}$ values of the fits done to calculate the veiling in the \ion{Li}{i}\,6707\,\AA\,absorption line.}
\label{table:laccandmacc}
\end{table*}

The corresponding value of the mass accretion rate is computed as:
\begin{equation}
\label{eq:Lacc}
    L_{\text{\,acc}} = \frac{G M_{\star} \dot{M}_{\text{\,acc}}}{R_{\star}} \left(1 - \frac{R_{\star}}{R_{in}}\right),
\end{equation}
\noindent where $R_{in}$ is the inner (truncation) radius. The truncation radius is assumed to be $R_{in}$ = 5\,$R_{\star}$~\citep{Gullbring1998}. Values of $M_\star$ and $R_\star$ are as in Table \ref{table:stellar_props}.

The $\dot{M}_{\text{\,acc}}$ ranges from a minimum of $\sim\!\num{3e-8} M_{\odot}\,$yr$^{-1}$ to a maximum of $\sim\!\num{7e-8} M_{\odot}\,$yr$^{-1}$. When averaged across the 15 nights, the result is  $\dot{M}_{\text{\,acc}}=\num{4.8e-8} M_{\odot}\,$yr$^{-1}$, in line with the values found in the literature. Small differences are due to slightly different distances ($140-160$\,pc) and masses ($0.5-1.1\,M_{\odot}$) adopted for RU\,Lup by different authors. This kind of variability aligns with what \citet[][Table 1]{Fischer2022} characterise as routine variability wherein there are day-to-day changes of $<1-2$\,mag in the optical and IR. This variability in CTTSs is described in \citet{Fischer2022} as a result of rotational, magnetospheric or inner-disk interactions with the star. In this study, we discuss the impact stellar rotation has (or does not have) on RU\,Lup's variability in \S\,\ref{stellar rotation}.

We want to stress here that while our values of $\dot{M}_{\text{\,acc}}$ depend on the stellar parameters adopted, the time variations of $\dot{M}_{\text{\,acc}}$, and those of $L_{\text{\,acc}}$, do not depend on them.

\subsection{Time-dependent veiling from metal lines} 
\label{subsection:veiling_metals}  

Additional information on the accretion variability of RU\,Lup can be obtained by measuring the excess emission in photospheric lines (veiling), as described in Eq. \ref{eq:veiling}. The excess emission in a given line is the sum of the continuum emission from the accretion shock at the line wavelength and of the emission of the same line in accreting matter \citep{Hartigan1989}. By measuring the amount of veiling in a CTTS, it is possible to quantify the amount of accretion activity occurring in the star during a given observation.

In \citet{Hartigan1989}, a quantitative method was developed to measure the amount of veiling present in a given CTTS. By comparing the spectrum of the accreting CTTS, with a template spectrum that represents only the photosphere (no accretion present), the dimensionless quantity, veiling, is produced:
\begin{equation}
    r_j = \frac{F_{excess}}{F_{photo}},
\label{eq:veiling}
\end{equation}
\noindent
where $r_j$ is the veiling in a specific wavelength interval $j$, $F_{excess}$ is the excess flux (due to accretion activity) and $F_{photo}$ is the photospheric flux, given from the template star. WTTS are often used for this purpose \citep[e.g.][]{Hartigan1991,Valenti1993,Johns-Krull1999,Dodin2012,McGinnis2018}.

The veiling present in a specific photospheric line can be similarly derived using the EWs of that line in the CTTS spectrum and the WTTS spectrum:
\begin{equation}
        r_j = \frac{EW_{photo}}{EW_{observed}} - 1,
\label{eq:ew_veiling}
\end{equation}
\noindent 
where EW$_{observed}$ is the equivalent width of the observed line in the CTTS spectrum and EW$_{photo}$ is the equivalent width of the same line in the template (WTTS) spectrum. Notably, the numerator in Eq. \ref{eq:ew_veiling} represents the total flux observed (the sum of the photospheric and excess fluxes), whereas the numerator in Eq. \ref{eq:veiling} represents only the excess flux.

Veiling, being the ratio of the excess flux over the photospheric flux, does not depend on extinction. Although it is not the case for RU\,Lup, this is very important, as the extinction is sometimes uncertain and it is known to vary with time in a large number of pre-main sequence stars \citep[e.g. the dippers studied by][]{ContrerasPena2017}.

Photospheric lines of different species and properties may exhibit different veiling if accretion-related line-emission is important, as proposed, for example, by \citet{Dodin2012}. 

\begin{figure}
    \centering
    \includegraphics[width=0.9\linewidth]{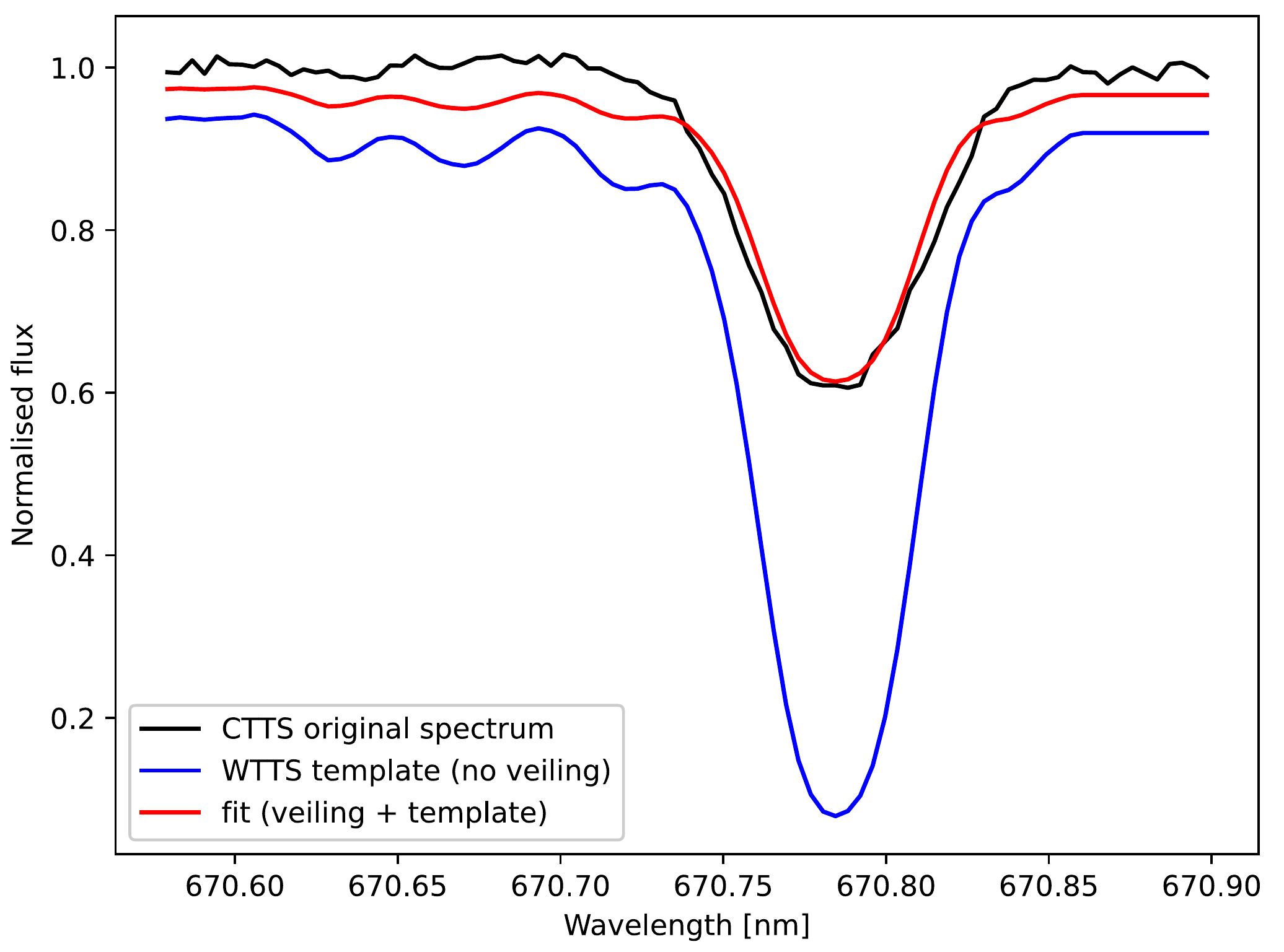}
    \caption{Example of fitting a photospheric (\ion{Li}{i}\,6707\AA) line in one of the RU\,Lup spectra (21-06-2011) with the WTTS template from TAP\,45 to estimate the veiling present in this region of the spectrum.}
    \label{fig:veil_fit}
\end{figure}

This study aims to add to the conversation concerning the distinction between continuum veiling, that is, the contribution to the measurement of veiling due to a continuum excess, and the contribution of specific line emission. We have, therefore, investigated veiling in individual lines in a clean region of our spectra to attain the most accurate measurements possible. This allows us to delve deeper into how veiling from metal lines relates to the accretion luminosity.

Therefore, we focus our analysis on three intrinsically deep metal lines in the region $5600-6700$\,\AA, namely the \ion{Li}{i}\,6707\AA, \ion{Mn}{i}\,6021\AA\,and \ion{Na}{i}\,5688\AA\,lines. The latter were chosen as they are within our wavelength range of interest and have a sufficiently high signal-to-noise ratio (S/N) for veiling measurements. 

The \ion{Li}{i}\,6707\,\AA\,line seen in absorption in young K7 stars has the deepest absorption in the optical wavelength range. It has an average EW$_{\ion{Li}{i}}= 0.17\pm0.03$\,\AA. We analysed this line by comparing it to a K7 WTTS and estimated the amount of veiling present as described in \S\,\ref{section:observations}. An example of our fits is provided in Fig.~\ref{fig:veil_fit}, and in Appendix \ref{appendix:calc_veiling} our method for calculating veiling is explained in detail. Preliminary analysis of the observed veiling produced an average over the 15 nights of $r_{\,\ion{Li}{i},\,\text{avg}}$(obs) = $1.92\pm0.03$. This observed value is affected by there being different abundances of Li in the template star and RU\,Lup, as discussed in \S\,\ref{section:LiVeiling}. The abundance-corrected average is $r_{\,\ion{Li}{i},\,\text{avg}}$ = $3.25\pm0.20$, which is used throughout this study. The nightly results are shown in column 8 of Table \ref{table:laccandmacc}.
  
Veiling measurements of the \ion{Na}{i}\,5688\,\AA\, and \ion{Mn}{i}\,6021\,\AA\, photospheric lines can be found in columns 5 and 6 of Table \ref{table:laccandmacc} and their averages over the 15 nights are $r_{\,\ion{Na}{i},\,\text{avg}}= 3.27\pm0.63$ and $r_{\,\ion{Mn}{i},\,\text{avg}}= 3.29\pm0.67$. As these photospheric lines do not have as high an S/N as that of the \ion{Li}{i}\,6707\,\AA, and are intrinsically weaker, there is higher uncertainty in the measured veiling. 

The additional photospheric metal lines \ion{Ti}{i}\,6261.10\,\AA, \ion{Ca}{i}\,6431.10\,\AA,\,and \ion{Ni}{i}\,6643.63\,\AA\,were initially considered as well; however, these lines were too veiled to get accurate measurements. This illustrates the difficulty in measuring the veiling in high accretors as there is a balance wherein the veiling is high enough to be noticeable in the spectrum, but is not so high that the photospheric lines are completely `washed out'.

\subsection{Wind \& jet lines}
\label{subsection:forbidden_lines} 

In the studied spectral range, there are multiple forbidden emission lines associated with outflowing material from young stars~\citep{Edwards1987, Hamann1994, Hartigan1995}. Present in these data with a strong S/N are the [\ion{O}{i}]\,6300\,\AA\,and [\ion{S}{ii}]\,6730\,\AA\,emission lines. The flux calibrated spectra of these lines from the 12 epochs can be seen in Fig.~\ref{fig:outflow_lines}. The vast majority of the emission from the [\ion{O}{i}]\,6300\,\AA\,line is blue-shifted and it is completely blue-shifted in the [\ion{S}{ii}]\,6730\,\AA. Both lines have a very similar profile with wings reaching very high systematic velocities, namely $\sim\!-300$ and $\sim\!-380$\,km\,s$^{-1}$ for the [\ion{S}{ii}]\,6730\,\AA\,and the [\ion{O}{i}]\,6300\,\AA\,lines, respectively. These wing velocity values are slightly higher than those found for RU\,Lup in the observations of \citet{Fang2018} and \citet{Banzatti2019} in 2008 ($\sim\!-270$\,km\,s$^{-1}$ for [\ion{O}{i}]\,6300\,\AA) and \citet{Whelan2021} in 2012 ($\sim\!-250$ and $\sim\!-280$\,km\,s$^{-1}$, for the [\ion{S}{ii}]\,6730\,\AA\,and the [\ion{O}{i}]\,6300\,\AA\,lines, respectively). All these values for the wings of the HVC in RU\,Lup are typically higher than those of other TTS with this forbidden emission, often by at least 100\,km\,s$^{-1}$, as can be seen in Figure~17 of \citet{Fang2018} for the case of the [\ion{O}{i}]\,6300\,\AA\,line.

These two line profiles show two distinct components: an HVC peaking at $\sim\!-200$ km s$^{-1}$ and a LVC peaking at $\sim\!-17$ km s$^{-1}$. The LVC of both lines is dominated by a narrow low-velocity component (NLVC), while the HVC contains two components, with one less defined very broad component at $\sim\!-75$ and $\sim\!-105$\,km s$^{-1}$ for the [\ion{S}{ii}]\,6730\,\AA\,and the [\ion{O}{i}]\,6300\,\AA\,line, respectively (see \ref{fig:outflow_line_fits}). The broad component of the [\ion{O}{i}]\,6300\,\AA\,line has a broader width (roughly 400\,km\,s$^{-1}$) than that of the [\ion{S}{ii}]\,6730\,\AA\,line (roughly 200\,km\,s$^{-1}$). These are quite common features in the jets and winds of CTTSs \citep{Rigliaco2013,Simon2016,McGinnis2018}.

While the main component of the HVC is tracing the collimated jet, the broad component is often associated with the base of the jet or a disk-wind and the NLVC with a slow-moving compact disk-wind or a photoevaporative wind from the disk~\citep[see e.g.][]{Ray07, Simon2016}. In the case of RU\,Lup, the different nature of such components has been recently confirmed by the spectro-astrometric analysis of \citet{Whelan2021}. Here the authors showed that the HVC and the broad component come from the extended jet and base of the jet, respectively, whereas the NLVC originates from a very compact ($\lesssim\,$8\,au) MHD disk-wind.

Interestingly, our measurements of the HVC and LVC radial velocities match very well the total jet and wind velocities measured or inferred in other CTTSs of similar mass~~\citep[see e.g.][ and references therein]{Ray07}. This is a further indication that the disk geometry is almost face-on~\citep[$16\degr\pm\,^{6\degr}_{8\degr}$;][]{GRAVITY}, namely the outflow is approaching the observer with a small inclination angle.

Indeed, because of the disk geometry, very little of the red-shifted emission from the outflow is observed, as it is located on the side of the star facing away from us, hidden by the disk. This explains why the red-shifted HVC is not detected in our spectra (see Figure~\ref{fig:outflow_lines}).

\section{Discussion}
\label{discussion}  

The results presented in Sec.~\ref{results} show that accretion in RU\,Lup varies from night-to-night over a period of 15 nights by about a factor of two. All the analysed accretion tracers (emission line luminosities, profiles, veiling) give consistent results. Notably, the average value of $\dot{M}_{\text{\,acc}}$ in this period agrees very well with values found in the literature at different epochs \citep{Alcala2017}. It is therefore likely that the variability pattern we found is typical of RU\,Lup's accretion at this stage of its evolution. 

In addition, as we mentioned in \S\,\ref{subsection:macc_var}, the extinction of RU\,Lup is negligible and has not been taken into consideration in this study. As the EW of each absorption line analysed changes slightly over our short timescales, this indicates that such variability is due to veiling and not to extinction.

\begin{figure}
    \centering
    \includegraphics[width=1.0\linewidth]{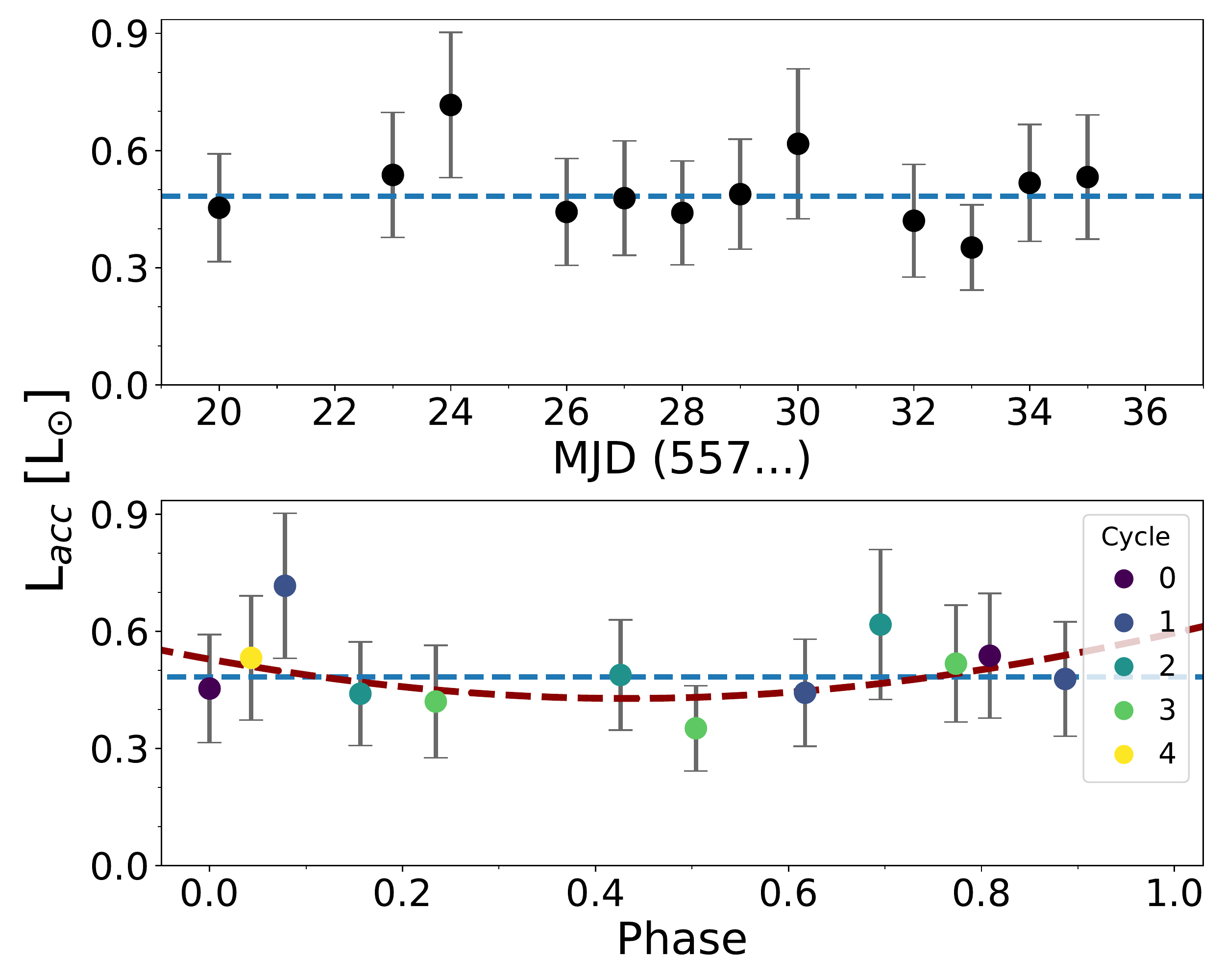}
    \caption{$L_{\text{\,acc}}$ variability across the 12 epochs (upper panel) and folded in phase with the rotation of the star (lower panel). The period is 3.71 days \citep{Stempels2007}. The colour of the data points in the lower panel indicates the cycle of rotation of the star. The parameters of the sinusoidal fit are: amplitude\,=\,$-0.4\pm0.1\,L_{\odot}$ and offset\,=$\,0.8\pm0.1\,L_{\odot}$. The blue dashed line indicates the median value of accretion luminosity of 0.48\,$L_{\odot}$.}
\label{fig:lacc_v_phase}
\end{figure}

As Fig. \ref{fig:outflow_lines} clearly shows, there is no variability of the line profile or flux of the [\ion{O}{i}]\,6300\,\AA\,or [\ion{S}{ii}]\,6730\,\AA\,forbidden lines in the timescale covered by our data. As was discussed in \S\,2.2, this is expected for the HVC, however if either of the other components were to originate from an MHD disk wind located in the innermost region of the disk, where accretion occurs, then we could expect to see some variability of that component on similar timescales as accretion. \citet{Simon2016} and \citet{McGinnis2018} show that the NLVC of the [\ion{O}{i}]\,6300\,\AA\,line originates from the inner disk, but far enough from the central object that it should not be directly impacted by accretion ($\,>\!0.5\,$au, which was confirmed for RU Lup by \citet{Whelan2021}), whereas the BC seems to trace a wind from closer in (between $\sim0.05\,$au and $\sim0.5\,$au from the central object). This latter component has velocities of $\lesssim\!50\,$km\,s$^{-1}$ and is therefore a LVC, unlike the BC of the [\ion{O}{i}]\,6300\,\AA\,line of RU\,Lup, which is more blue-shifted. It is therefore evident that the broad component seen in the [\ion{O}{i}]\,6300\,\AA\,line profile of RU Lup is not the same component as the broad low-velocity component (BLVC) described by \citet{Simon2016} and \citet{McGinnis2018} in their samples, but is likely an intermediate velocity component linked to the base of the jet \citep{Whelan2021}. This star therefore lacks the BLVC which is linked to an MHD disk wind in the innermost part of the disk.

\subsection{$L_{\text{\,acc}}$ relation to the stellar rotation}
\label{stellar rotation}

Variation in the observed $L_{\text{\,acc}}$ is sometimes associated with the stellar rotation if the line of sight intercepts a single or multiple accretion spots \citep{Costigan2014}, the properties of which are relatively stable over several rotation periods. For instance, \citet{Alencar2018} show that the veiling measured in LkCa15 varies by a factor of $\sim\!5$ over the course of a few rotation cycles, and that this variability folds well in phase with the star’s rotation period (see their Fig. 5). Similarly, \citet{McGinnis2020} show that the veiling variability and, even more pronounced, the flux of the {\ion{He}{i}} emission line at 5876\,\AA\  of the star IP Tau have a clear modulation at the star’s rotation period.

\begin{figure}
    \centering
    \includegraphics[width=1.0\linewidth]{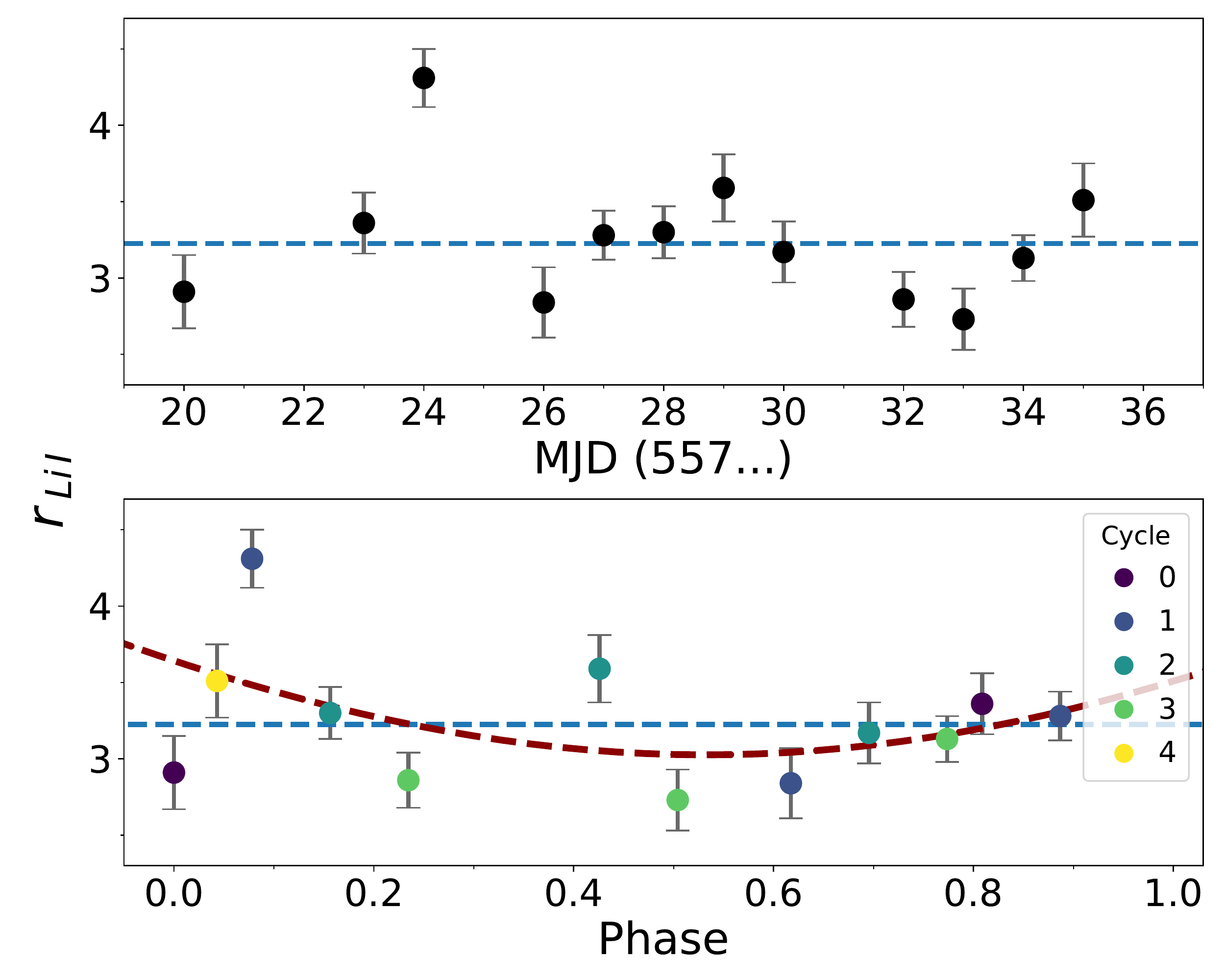}
    \caption{Veiling in the \ion{Li}{i}\,6707\,\AA\,line. $r_{\,\ion{Li}{i}}$ variability across the 12 epochs (upper panel) and folded in phase with the rotation of the star (lower panel). The period is 3.71 days \citep{Stempels2007}. The colour of the data points in the lower panel indicate the cycle of rotation of the star. The parameters of the sinusoidal fit are: amplitude\,=\,$1.6\pm0.1$ and offset\,=\,$4.7\pm0.1$.The blue dashed line indicates the median value of \ion{Li}{i} veiling of 3.23.}
\label{fig:veiling_date}
\end{figure}

However, given the nearly face-on geometry of RU\,Lup's disk~\citep{Gahm2013,GRAVITY}, we do not expect that the observed variability is related to stellar rotation. To confirm this hypothesis, we study how $L_{\text{\,acc}}$ varies as a function of the stellar rotation period. In Fig.\,\ref{fig:lacc_v_phase}, the variability of $L_{\text{\,acc}}$ is presented across the roughly two weeks of the observations (upper panel) and folded in phase with the rotation of the star (lower panel) which has a period of 3.71 days \citep{Stempels2007}. To quantify the amount of $L_{\text{\,acc}}$ variability potentially due to rotation, we fit a sinusoidal curve to the folded data. The resulting fit (red-dashed curve) is shown in the lower panel of Fig.\,\ref{fig:lacc_v_phase} and the fit parameters are listed in the figure's caption. Similar plots are shown in Fig.\,\ref{fig:veiling_date} for the veiling derived from the \ion{Li}{i}\,6707\,\AA\,line ($r_{\,\ion{Li}{i}}$) variability $vs.$ time and folded in phase with stellar rotation (upper and lower panel, respectively). 

If the variability seen in these two quantities was solely due to RU\,Lup's rotation, the lower panels of Fig.\,\ref{fig:lacc_v_phase} and Fig. \ref{fig:veiling_date} would show a distinct sinusoidal trend. However, the data vary in a more complex way, indicating that an intrinsic variability in the star's accretion exists. Indeed, the curve amplitude in the fit of Figure~\,\ref{fig:lacc_v_phase} is only $L_{\text{\,acc}}=0.04$\,L$_\sun$ implying that, if present, only 7\% of the $L_{\text{\,acc}}$ variability can be attributed to stellar rotation. 

\subsection{Intrinsic accretion variability}
\label{intrinsic_acc_var}

As the bulk of RU\,Lup's accretion variability cannot be due to stellar rotation nor to extinction, it must then be related to variations of the mass accretion rate (a factor of 2 or 0.31\,dex, see Table~\ref{table:laccandmacc}) onto the star during our 15-day spectroscopic monitoring. On our observational timescale, such variability is similar to what was found in other photometric and spectroscopic studies of CTTSs \citep{Costigan2012,Venuti2014,Zsidi2022}. 

By spectroscopically monitoring a sample of 15 CTTSs and Herbig Ae/Be stars, \citet{Costigan2014} derived that their mean accretion rates change from 0.04 to 0.4\,dex on a daily basis. The photometric monitoring of $\sim$750 young stars by \cite{Venuti2014} found a 0.5~dex variability over a period of two weeks.

For what concerns RU\,Lup's accretion variability, long-term photometric monitoring by \citet{Siwak2016} already showed stochastic variability and lack of a single stable periodic pattern in its light-curves. The authors inferred that an unstable accretion regime, typical of CTTSs with high-mass accretion rates, operates in RU\,Lup. In particular, they argue that RU\,Lup's photometric variability is due to increased mass accretion rates that can prompt Rayleigh–Taylor (RT) instabilities, moving accretion from a stable to an unstable regime, as predicted by several 3D MHD simulations of disk accretion onto a rotating
magnetised star \citep[see][]{Romanova2008,Kulkarni2008,Kulkarni2009}. 

Although several ingredients can influence such a transition (strength of the magnetic field - $B$, stellar rotation rate, misalignment between the star's rotation axis and its magnetic poles), \cite{Romanova2008} showed that the main player in the development of RT instabilities is the mass accretion rate.

During the stable regime, the accretion disk is blocked by the star’s magnetic field, and, at the magnetospheric radius ($\sim$5\,R$_*$), the accreting matter flows around the magnetosphere, forming two funnels that deposit matter near the star’s magnetic poles, making accretion relatively stable. An increase in accretion rates and thus in density squeezes the magnetosphere in the equatorial region. As a consequence, it is energetically impossible for the inner disk matter to follow the resulting magnetospheric field lines. RT instabilities develop, causing matter to move towards the star across the field lines, until it reaches field lines that are energetically possible to follow. The matter is wound along these lines, forming miniature funnel flows and accreting at lower latitudes \citep{Romanova2008,Kulkarni2008,Kulkarni2009}.

We can compare our $\dot{M}_{\text{\,acc}}$ results with the predictions from \cite{Romanova2008,Kulkarni2008,Kulkarni2009} and verify whether RT instabilities might be the source of the observed variability in RU\,Lup. The $\dot{M}_{\text{\,acc}}$ boundary value between the stable and unstable regime in a CTTS is provided by Eq.~1 of \citet{Kulkarni2008} and depends on the $B$, $M_{\star}$ and $R_{\star}$. For our calculations, we use the stellar parameters reported in Table~\ref{table:stellar_props}. The strength of RU\,Lup's magnetic field is not well constrained in the literature; however, \citet{Johnstone1986} provide an upper limit of $\sim$500\,G, which we adopt as $B$ of RU\,Lup. Eq.~1 of \citet{Kulkarni2008} with RU\,Lup's parameters gives $\dot{M}_{\text{\,acc}}\!\sim1.6\,\times\,10^{-8}\,M_\odot\,$yr$^{-1}$ as the critical mass accretion rate whereby the stable regime shifts to an unstable one. This $\dot{M}_{\text{\,acc}}$ is lower than the observed range of $\dot{M}_{\text{\,acc}}$ values in our target. 
Notably, lower $B$ values would provide a lower boundary value for $\dot{M}_{\text{\,acc}}$. We can thus confirm the possibility an of unstable accretion regime in RU\,Lup.

\citet{Kurosawa2013} employed the results of the 3D simulations of \cite{Romanova2008} to compute time-dependent accretion rates and calculate Hydrogen line profiles for a CTTS slightly more massive (0.8\,M$_\odot$) and more magnetised (1000\,G) than RU\,Lup, and with a slightly larger stellar radius (2\,R$_\odot$). As expected from Eq.~1 of \citet{Kulkarni2008}, the accretion rates derived by \citet{Kurosawa2013} for such a CTTS in the unstable regime are slightly higher than ours~\citep[see lower panel of Fig.~2 of][]{Kurosawa2013}. However, the relative $\dot{M}_{\text{\,acc}}$ variations are comparable to the observed variations in RU\,Lup.

\subsection{Veiling in the \ion{Li}{i}\,6707\,\AA\,line}
\label{section:LiVeiling}

The \ion{Li}{i}\,6707\,\AA\,photospheric line is the deepest absorption line in our spectra and benefits from a high S/N, making it a more enticing photospheric line with which to study the veiling. However, this use of the \ion{Li}{i}\,6707\,\AA\,line potentially suffers from the uncertain Li abundance ($A$(Li) = log\,$N$(Li)) of pre-main sequence stars, due to the onset of Li nuclear burning. The Li abundance decreases as a star gets older, in a way that depends on its stellar properties, that is, mass and temperature \citep{Palla2007}.

It has been hypothesised that WTTSs are simply more evolved CTTSs with dissipated disks \citep{Walter1988,Bertout1989}. However, there are CTTSs and WTTSs that coexist \citep[e.g. Fig. 15 of][]{Kenyon1995}, and their disk evolution is informed by their star-forming region and circumstellar environment \citep{Bonnell2007,Rosotti2014}. The details of this complex evolution are beyond the scope of this study, however, the implication is that the amount of Li depletion (a known indicator of stellar age) is often higher in WTTSs than CTTSs, but cannot be assumed \citep{Galli2015}.

Therefore, the template used to measure the veiling needs to take the Li abundance of the target CTTS and the template WTTS into account. For example, if the Li in the template star is depleted, there will be less Li in the photosphere to absorb light and the EW of the \ion{Li}{i} line will be less than an un-depleted star. Because the flux of the \ion{Li}{i}\,6707\,\AA\,line in the WTTS is compared with that of the CTTS to evaluate veiling, the abundances of Li must be similar, if not equal, ($A$(Li)$_{\text{WTTS}}\!\simeq A$(Li)$_{\text{CTTS}}$) to accurately identify the difference in line depth as a veiling effect.

\begin{figure}
    \centering
    \includegraphics[width=1.0\linewidth]{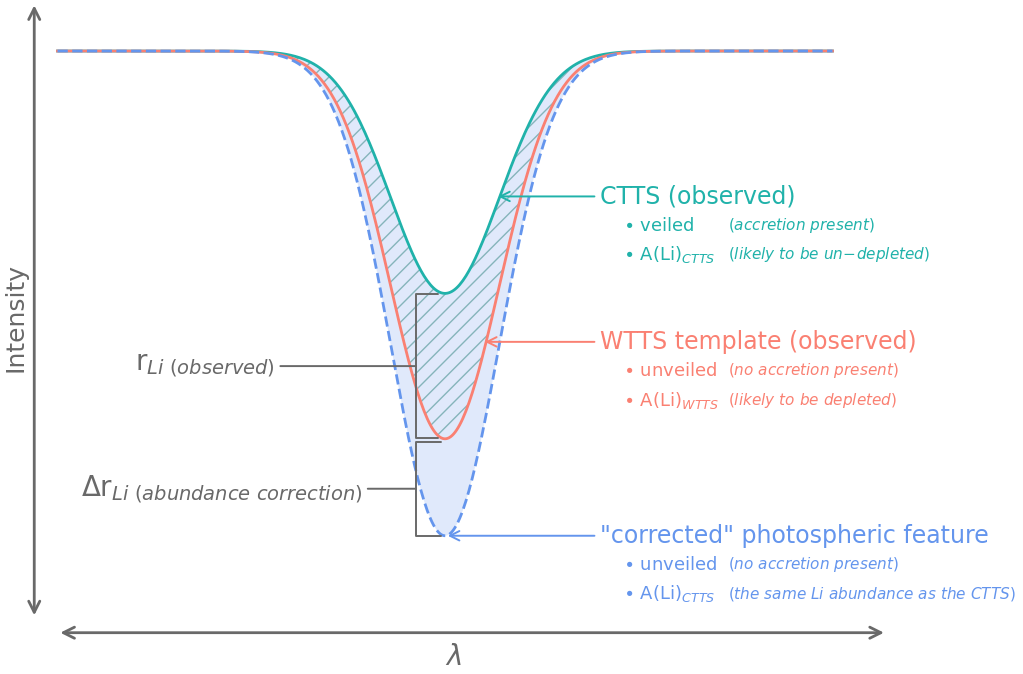}
    \caption{Diagram showing the difference in an observed CTTS (teal) \ion{Li}{i}\,6707\,\AA\,line and that of a WTTS (orange). The deepest absorption feature (dashed light blue) represents a spectrum with no veiling and the same abundance of \ion{Li}{i}\,6707\,\AA\,as the CTTS. The hatched region represents the difference in absorption depth between the CTTS and WTTS. The shaded region represents the difference in absorption depth between the CTTS and the unveiled CTTS spectrum.}
    \label{fig:corrected_abund_diagram}
\end{figure}

RU\,Lup has had its abundance measured as $A$(Li)\,>\,3.81 dex, which indicates no depletion of Li \citep{Biazzo2017}. On the other hand, TAP\,45 has $A$(Li) values ranging from 2.92 to 3.27, depending on the method used for determining it \citep[see Column 8 of Table~1 in ][their source \#51]{Sestito2008}. This means that our template has an $A$(Li) from 0.5 to 0.9 dex \textit{lower} than RU\,Lup, leading to an underestimate of $r_{\,\ion{Li}{i}}$. 

In order to account for the difference in Li abundance between our WTTS template and RU\,Lup, we derive a correction factor by comparing the theoretical unveiled EW$_{\text{Li}}$ of RU\,Lup \citep{Biazzo2017} to the EW$_{\text{Li}}$ that we measured from our template star, TAP\,45. \citet{Biazzo2017} analyse a sample of YSOs, which includes RU~Lup, that were observed with X-Shooter \citep[this sample is described in][]{Alcala2017,Frasca2017}. They measure the EW$_{\text{Li}}$ for each star from the observed data and then unveil these equivalent widths using the values of $\!r_{6700\,\AA}$ estimated in \citet{Frasca2017} from the code ROTFIT \citep[e.g.][]{Frasca2006,Frasca2015}. 
This value of veiling is extrapolated between the values \citet{Frasca2017} found for $r_{6200\,\AA}$ and $r_{7100\,\AA}$. The notation EW$_{\text{Li,\,CTTS}}^{\text{corr.}}$ represents the equivalent width of the \ion{Li}{i}\,6707\,\AA\,line in a CTTS with the veiling effect removed (or `unveiled') \citep[see \S\,3.1.1 of][]{Biazzo2017}. The value found for RU\,Lup in \citet{Biazzo2017} is EW$_{\text{Li,\,RU\,Lup}}^{\text{corr.}}=988\pm10\,$m\AA.

This is illustrated in Fig. \ref{fig:corrected_abund_diagram}, where the teal line represents the observed \ion{Li}{i}\,6707\,\AA\,line of RU\,Lup, the orange line represents the template star, TAP\,45, and the dashed light blue line represents the absorption line of RU\,Lup when unveiled. The difference between the orange line's EW and the light blue line's EW is the difference of veiling measurement between the two templates. We call this the abundance correction factor ($\Delta\,r_{\text{Li}}$) as it represents the discrepancy between the WTTS template and the template EW$_{\text{Li}}$ of RU\,Lup unveiled: a difference only in Li abundance.

Rewriting Eq. \ref{eq:ew_veiling} using this notation, the correction factor is defined as:

\begin{equation}
\label{eq:r_corr}
    \Delta\,r_{\text{Li}} = \frac{EW^{\,\text{corr.}}_{\text{Li}}}{EW_{\text{Li}}} -1,
\end{equation}

\noindent
where EW$_{\text{Li}}^{\text{corr}}=988\pm10\,$m\AA\,and EW$_{\text{Li}}=352\pm10\,$m\,\AA\,(the template star's (TAP\,45) equivalent width of the \ion{Li}{i}\,6707\,\AA\,line.) This gives $\Delta\,r_{\,\text{Li}}$ = 1.8. A resultant total veiling can be found by summing $\Delta\,r_{\,\text{Li}}$ and $r_{\,\text{Li}}$(obs), which is taking the differing abundance between RU\,Lup and the template star TAP\,45 into account.  

\begin{figure}
    \centering
    \includegraphics[width=1.0\linewidth]{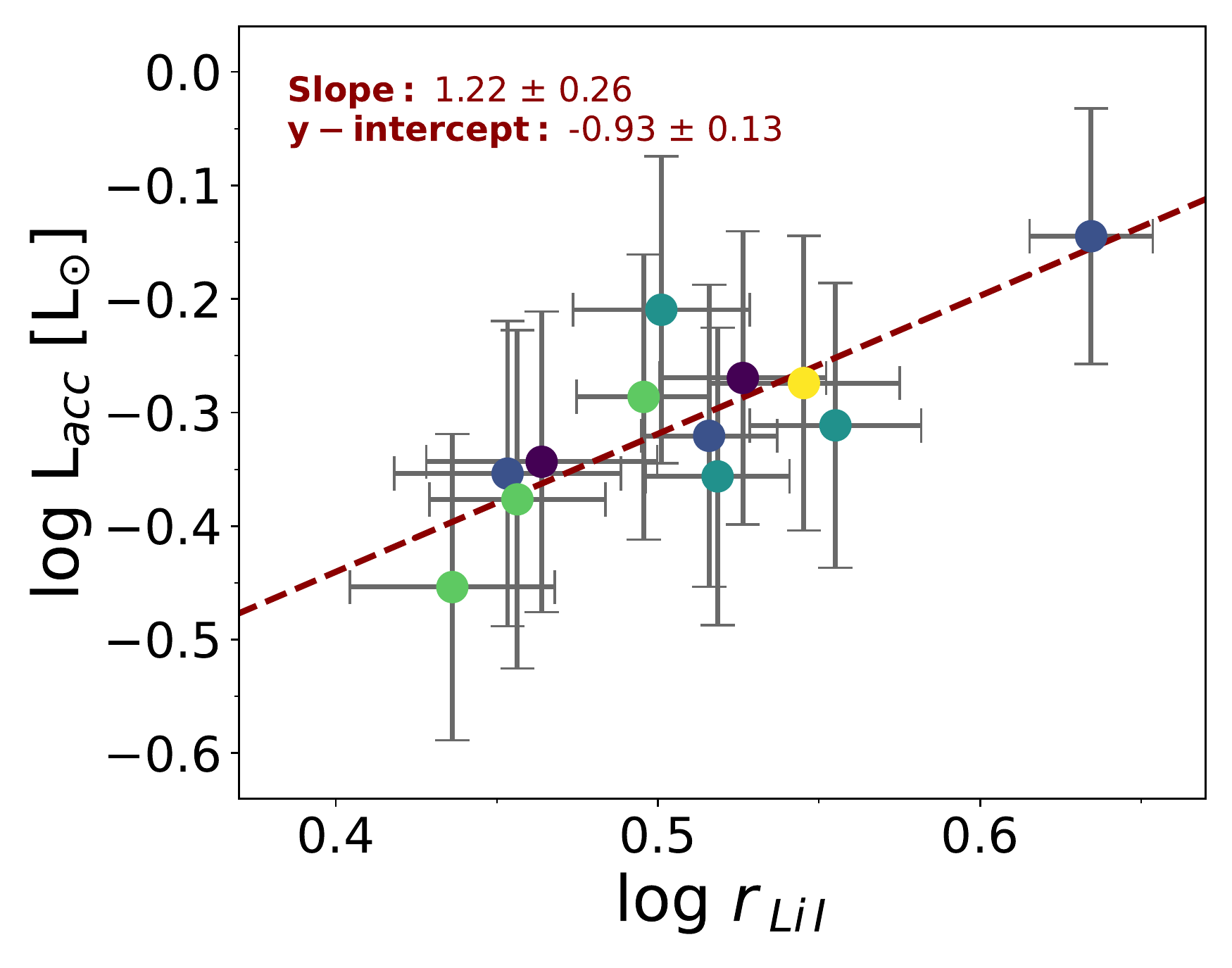}
    \caption{Fit representing how the veiling, estimated from the \ion{Li}{i}\,6707\,\AA\,absorption line (including a Li abundance correction - see \S\,\ref{section:LiVeiling}), correlates with the $L_{\text{\,acc}}$ logarithmically. The colours of the dots correspond to the same colour scheme used throughout this paper, as shown in Fig \ref{fig:outflow_lines}.}
    \label{fig:LOG_veil_v_lacc}
\end{figure}

It should be noted that there are large uncertainties in this correction factor, because of the way that the unveiled EW$_{\text{Li}}$ of RU~Lup was determined. RU\,Lup is a strongly veiled CTTS in which continuum veiling is likely not the dominant form of veiling \citep{Gahm2008}, meaning that individual photospheric absorption lines suffer from different amounts of veiling due to the filling in of lines by line emission (this will be explained further in the next section \S~\ref{subsection:cont_line_veil}). Therefore using an extrapolation of the veiling from other photospheric lines is not very accurate to determine the amount of veiling in the {\ion{Li}{i}} line. 
However, the similarities between the spectral types, $v$\,sin\,$i$, and other stellar properties of RU\,Lup and the template make the corrected measurements good approximations. In any case, this effect does not affect the relative variation observed in the veiling of the {\ion{Li}{i}} line and the results shown in Fig. \ref{fig:LOG_veil_v_lacc}.

Another method to determine veiling in the {\ion{Li}{i}} line would be to simply compare the EWs of the {\ion{Li}{i}} line in our observed spectra with the EW$_{\text{Li}}$ given by \citet{Biazzo2017} using Eq. \ref{eq:ew_veiling}. However, we chose instead to use the same method as was used to determine veiling for the \ion{Na}{i} and \ion{Mn}{i} lines and calculate a correction factor for the ${\ion{Li}{i}}$ abundance to maintain self-consistency. Regarding the measurements of these other two metal lines, it is preferable to compare the observed lines with a WTTS spectrum to determine veiling, rather than use a theoretical unveiled EW of each line in Eq.~\ref{eq:ew_veiling} as was done for {\ion{Li}{i}} by \citet{Biazzo2017}. There should not be a measurable difference in the abundance of these elements between our template and RU\,Lup, and therefore extrapolating values of veiling found from several photospheric lines by \citet{Frasca2017} to get an unveiled EW would only insert additional uncertainties to our measurements.

Even given these caveats, the most interesting finding of this study stands, the relative variation of the veiling with time does not depend on the uncertainties of the \ion{Li}{i}\,6707\,\AA\,abundances in RU\,Lup itself nor in the template.

Fig. \ref{fig:LOG_veil_v_lacc} shows our result that there is a close proportionality between $r_{\ion{Li}{i}}$ and $L_{\text{\,acc}}$ for RU\,Lup. The absolute value of the accretion luminosity for a given value of $r_{\ion{Li}{i}}$ may depend on a number of stellar and accretion properties, and the possibility to derive $L_{\text{\,acc}}$ from the \ion{Li}{i}\,6707\,\AA\,veiling alone needs to be explored further. However, our results suggest that in any individual star the variation of $r_{\ion{Li}{i}}$ may be correlated with variabilities in $L_{\text{\,acc}}$.

\subsection{Continuum and line veiling}
\label{subsection:cont_line_veil}

\begin{figure*}
    \centering
    \includegraphics[width=0.33\linewidth]{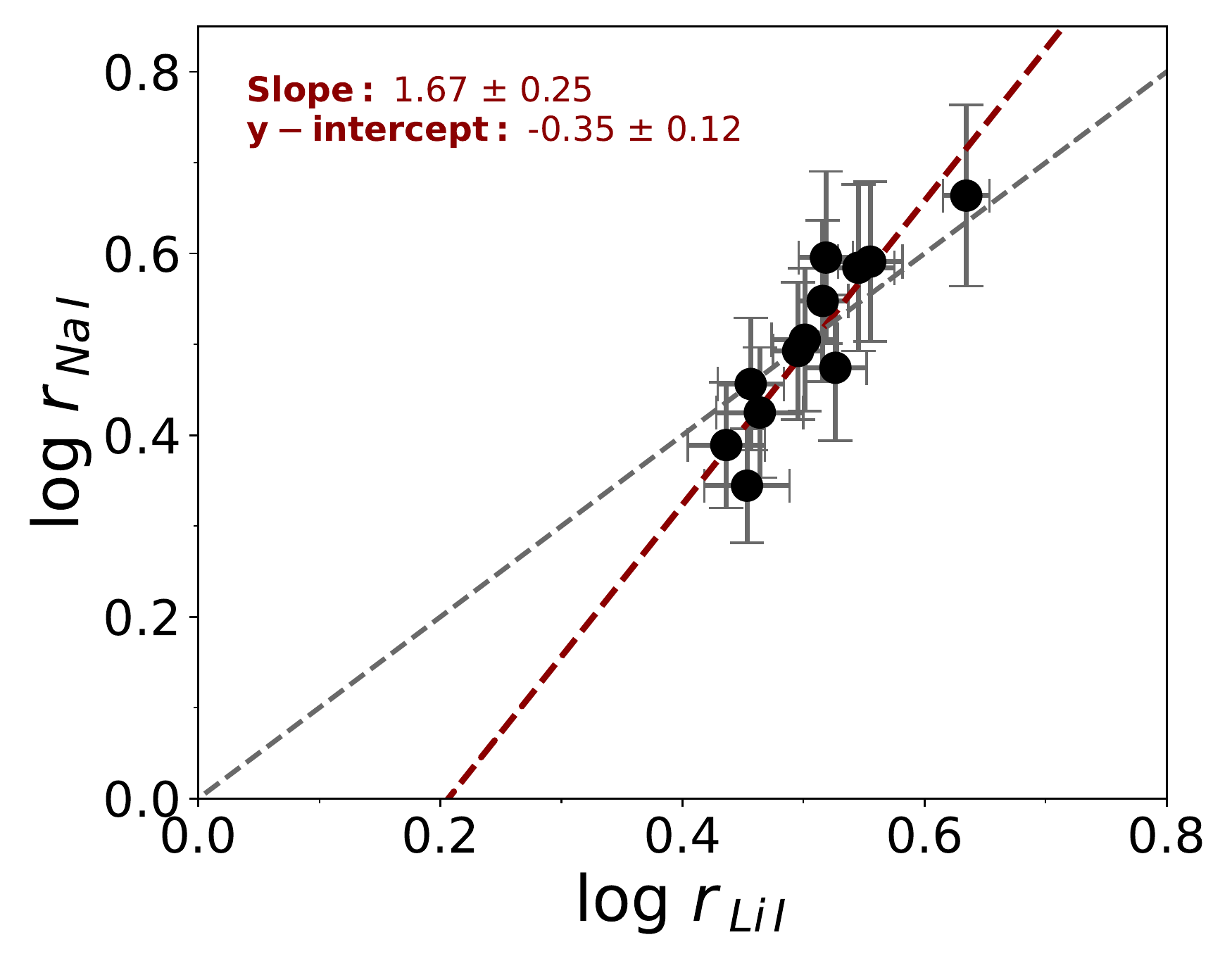}
    \includegraphics[width=0.33\linewidth]{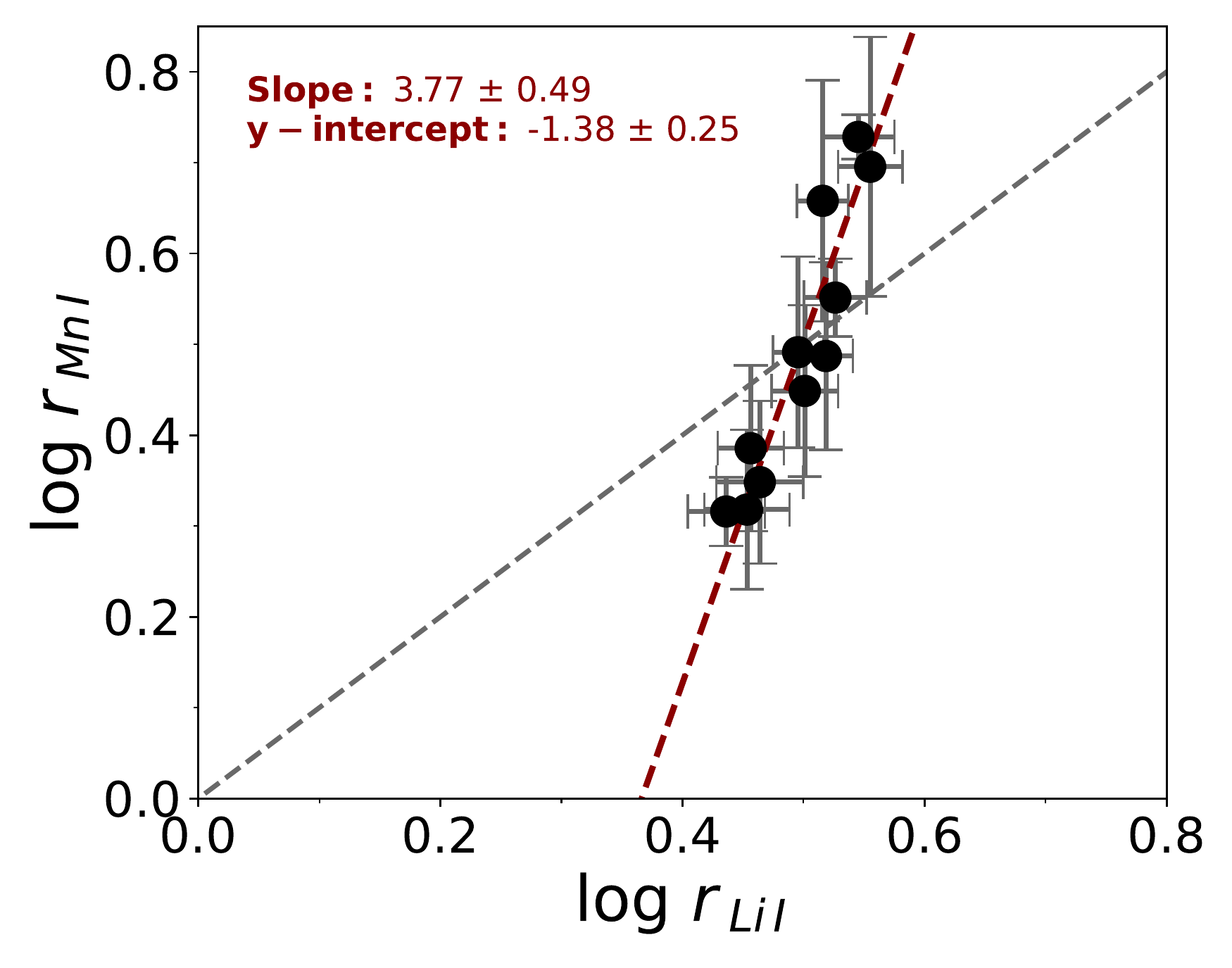}
    \includegraphics[width=0.33\linewidth]{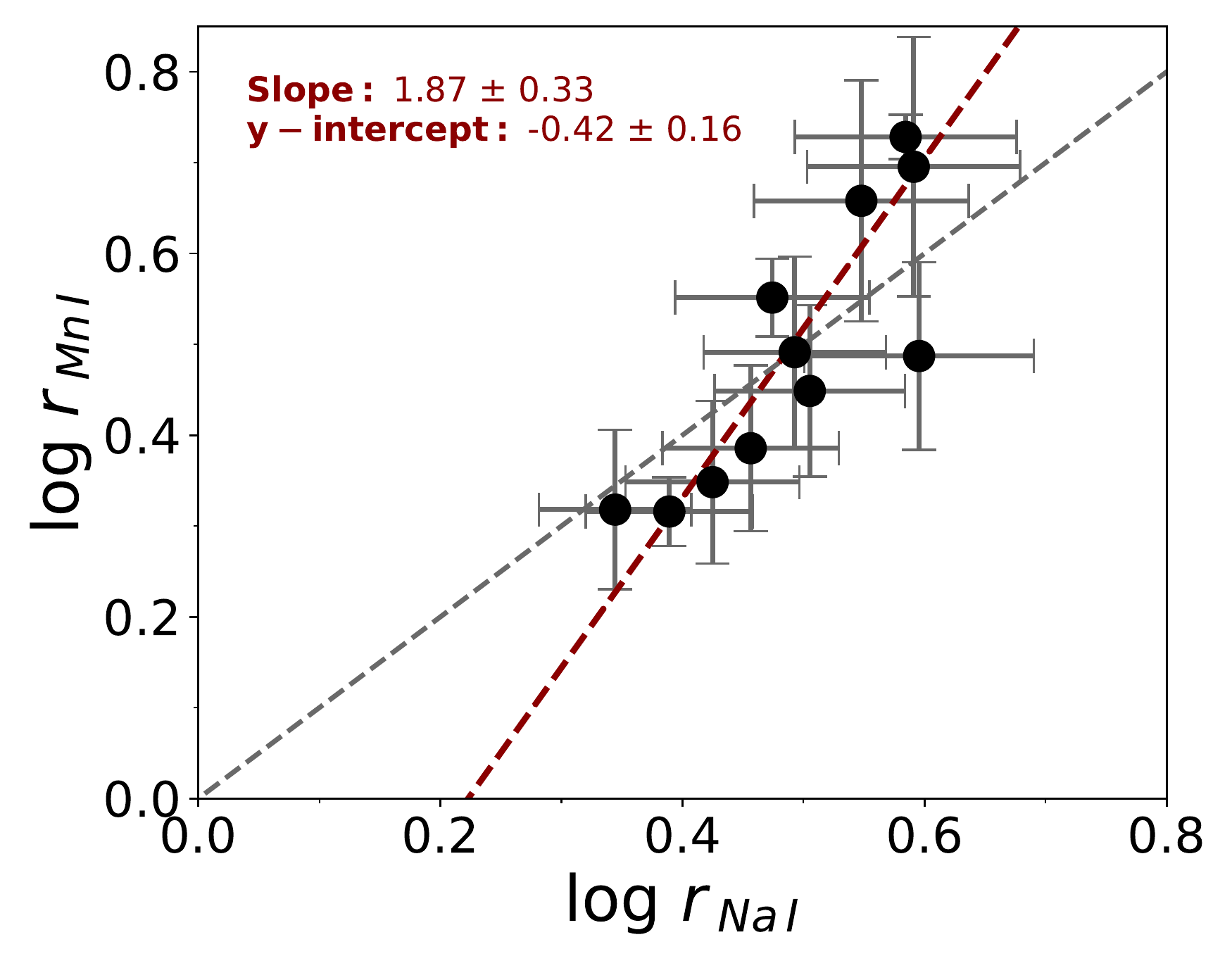}
    \caption{Log plots of the veiling measured in the metal lines of \ion{Li}{i}\,6707\,\AA, \ion{Na}{i}\,5688\,\AA\,and \ion{Mn}{i}\,6021\,\AA\,in comparison with each other. These plots allow us to visualise how the veiling can vary from one line species to another and that this is indicative of line emission veiling contributing to the total veiling measured in each.}
    \label{fig:veil_v_veil}
\end{figure*}

As noted in the previous section, the veiling values of the three metal lines we have studied roughly agree within the uncertainties. However, a closer inspection shows that, while there is a rather tight correlation,  the values of the veiling of different lines are not proportional to each other, as shown in Fig.8. In a system with only continuum excess emission, the veiling in these lines, which spreads over a relatively small interval of wavelengths,  should present a 1:1 relationship with one another, as we expect that the continuum will change by a very similar amount over this interval. This, however, is not going to be the case if accretion-related emission in the metal lines is also contributing to the overall veiling values.

While the \ion{Na}{i}\,5688\,\AA\, line roughly traces that of the \ion{Li}{i}\,6707\,\AA, the \ion{Mn}{i}\,6021\,\AA\, veiling increases sharply as the \ion{Li}{i}\,6707\,\AA\, veiling increases. This implies that the \ion{Mn}{i}\,6021\,\AA\, line emission increases more strongly with an increase in $L_{\text{\,acc}}$ than the \ion{Li}{i}\,6707\,\AA\, and \ion{Na}{i}\,5688\,\AA\, line emission. However, \ion{Li}{i}\,6707\,\AA, \ion{Na}{i}\,5688\,\AA,\, and \ion{Mn}{i}\,6021\,\AA\, all show strong reactions to an increase in accretion in RU\,Lup. 

This accretion-related emission in metal lines, `line veiling', has been observed in a number of accreting TTS \citep[e.g.][]{Bertout1984,Hartigan1989,Gahm2008}. In some, the line veiling has been found to be dominant, among them RU\,Lup \citep{Gahm2008,Gahm2013}. The authors compare the $V$-band photometric variability of the star with the variability of the veiling and find that the veiling is not correlated, or only weakly related, to the stellar brightness during states of high veiling. They argue that there are two sources of veiling: a continuous excess emission, and narrow emission lines filling-in the photospheric absorption lines, both related to the accretion shocks and foot-prints of the accretion funnels. Line emission becomes the dominant component for high-veiling states ($r>2$). In this case, a change in veiling does not lead to a corresponding change in stellar brightness, as it would happen if the veiling was caused by continuous emission alone.

Indeed this lack of correlation in highly veiled, and thus highly accreting CTTSs, is confirmed by \citet{Rei2018}. For a small sample of young active stars, \citet{Rei2018} find that the veiling is strongly line-dependent and is larger in strong photospheric lines and weaker or absent in the weakest ones. This concurs with what we have found in the three metal lines, namely, the stronger lines exhibit stronger reactions to increases in accretion.

\section{Summary and conclusions}
We have investigated the accretion variability of RU\,Lup over time through the \ion{H}{i} line luminosities and the veiling present in the photospheric metal lines \ion{Li}{i}\,6707\,\AA, \ion{Na}{i}\,5688\,\AA, and \ion{Mn}{i}\,6021\,\AA. Our main results are the following:

\begin{itemize}
    \item  $L_{\text{\,acc}}$ and $\dot{M}_{\text{\,acc}}$ were observed to vary by up to a factor of $\sim\!2$ over the 15 nights. This variability is an intrinsic variation of the accretion rates. In particular, it does not reflect variations related to the stellar rotation.
    \item The average value of the mass accretion rate found is $\dot{M}_{\text{\,acc,\,avg}}=\num{4.8e-8}\,M_{\odot}\,$yr$^{-1}$. This is within the spread of values found in \citet{Herczeg2008} ($\num{1.8e-8}\,M_{\odot}\,$yr$^{-1}$) and \citet{Alcala2017} ($\num{6.7e-8}\,M_{\odot}\,$yr$^{-1}$).
    \item The wind-tracing emission lines (\,[\ion{O}{i}]\,6300\,\AA\, and [\ion{S}{ii}]\,6730\,\AA) are very stable over the time interval of these observations. They do not reflect the short-term changes in the mass accretion rate.
    \item In comparing the models of \citet{Romanova2008}, \citet{Kulkarni2008}, and \citet{Kulkarni2009}, we establish RU\,Lup is likely to be in an unstable accretion regime.
    \item The \ion{Li}{i}\,6707\,\AA\, absorption line is measured to have an average EW$_{\ion{Li}{i}}= 0.17\pm0.03$\,\AA\, and an average $r_{\,\ion{Li}{i}}= 3.25\pm\,0.20$ when the Li depletion in the template star TAP\,45 is taken into account.
    \item The temporal variability pattern over the 15 nights for $r_{\,\ion{Li}{i}}$ is strongly correlated to that of $L_{\text{\,acc}}$ (see Fig. \ref{fig:LOG_veil_v_lacc}). 
    \item The veiling measured in the photospheric absorption lines \ion{Na}{i}\,5688\,\AA\, ($r_{\,\ion{Na}{i},\,\text{avg}}= 3.27\pm0.63$) and \ion{Mn}{i}\,6021\,\AA\, ($r_{\,\ion{Mn}{i},\,\text{avg}}= 3.29\pm0.67$) demonstrate positive, although different, correlations with the veiling measured in \ion{Li}{i}\,6707\,\AA\, and $r_{\,\ion{Li}{i},\,\text{avg}}= 3.25\pm0.20$. This reveals the relative importance of emission line veiling in each individual line's total veiling. The veiling found in \ion{Mn}{i}\,6021\,\AA, for example, increases more rapidly than the other metal lines, given an increase in accretion. 
    \item The correlation we detect between changes in the accretion luminosity and changes in the veiling is strong and provides further evidence that the variability observed is not caused by extinction (the veiling measurements are unaffected by extinction because they are obtained by comparing two normalised spectra).
\end{itemize}
    
Because of the strength of the \ion{Li}{i}\,6707\,\AA\,line in CTTSs, its veiling can be an excellent tool in monitoring accretion variability that is independent of extinction and flux calibration. This relationship between variabilities in $r_{\,\ion{Li}{i}}$ and $L_{\text{\,acc}}$ over short timescales can be further explored in other CTTSs to characterise accretion activity regardless of their visual extinction.

A future multi-object study aimed at calibrating this relationship could benefit from the careful selection of template stars that have similar Li abundances as the targets. The correlation between these two quantities found in this paper need not only be applied to the \ion{Li}{i} photospheric line; however, its prevalence in TTSs and high S/N make it a better tool than other metal lines that are not always visible. The techniques explained in this paper also provide useful methods to calculate veiling in the absence of a perfect template star, opening the door for more research into the accretion processes in young stars.


\begin{acknowledgements}
\\
The authors would like to thank Juan Alcalá and J\'er\^ome Bouvier for insightful discussions on the topic. This project received funding from the European Research Council under Advanced Grant No. 743029, Ejection, Accretion Structures in YSOs (EASY). \\
\end{acknowledgements}

\bibliographystyle{aa} 
\bibliography{bib_cstock_44315.bib} 

\begin{appendix}
\label{appendix}

\onecolumn
\section{Integrated fluxes of accretion lines}
\label{appendix:fluxes}

Below are the tables containing the integrated fluxes of the accretion-tracing lines.

\begin{table}[h]
\centering
\begin{tabular}{l c c c c c}
\hline\hline  
\noalign{\smallskip}
Date & $F_{H\alpha}$ & $EW_{H\alpha}$ & $F_{H\beta}$ & $EW_{H\beta}$ \\ 
\noalign{\smallskip}
yyyy-mm-dd & $\times10^{-12}$ [erg s$^{-1}$ cm$^{-2}$] & [\AA] & $\times10^{-12}$ [erg s$^{-1}$ cm$^{-2}$] & [\AA] \\
\noalign{\smallskip}\hline
\noalign{\smallskip}
2011-06-08 & $22.96 \pm 0.02$ & $-142.22 \pm 0.03$ & $4.25 \pm 0.04$ & $-56.69 \pm 0.04$ \\
2011-06-11 & $25.45 \pm 0.01$ & $-131.85 \pm 0.03$ & $5.07 \pm 0.07$ & $-55.55 \pm 0.04$ \\ 
2011-06-12 & $30.35 \pm 0.02$ & $-119.31 \pm 0.03$ & $6.53 \pm 0.02$ & $-51.85 \pm 0.05$ \\
2011-06-14 & $22.50 \pm 0.01$ & $-133.94 \pm 0.03$ & $4.19 \pm 0.05$ & $-58.15 \pm 0.04$ \\
2011-06-15 & $24.10 \pm 0.01$ & $-125.81 \pm 0.03$ & $4.44 \pm 0.05$ & $-48.78 \pm 0.04$ \\
2011-06-16 & $21.96 \pm 0.01$ & $-117.41 \pm 0.03$ & $4.19 \pm 0.05$ & $-44.08 \pm 0.04$ \\
2011-06-17 & $23.03 \pm 0.01$ & $-111.94 \pm 0.03$ & $4.66 \pm 0.10$ & $-47.94 \pm 0.05$ \\
2011-06-18 & $30.12 \pm 0.01$ & $-149.78 \pm 0.03$ & $5.58 \pm 0.08$ & $-66.94 \pm 0.04$ \\
2011-06-20 & $22.44 \pm 0.01$ & $-134.46 \pm 0.03$ & $4.00 \pm 0.02$ & $-47.61 \pm 0.04$ \\
2011-06-21 & $18.34 \pm 0.01$ & $-114.83 \pm 0.03$ & $3.44 \pm 0.03$ & $-46.26 \pm 0.04$ \\ 
2011-06-22 & $24.61 \pm 0.01$ & $-125.28 \pm 0.03$ & $4.86 \pm 0.07$ & $-53.06 \pm 0.04$ \\
2011-06-23 & $25.92 \pm 0.01$ & $-127.09 \pm 0.03$ & $4.91 \pm 0.06$ & $-47.69 \pm 0.04$ \\
\noalign{\smallskip}
\hline \hline
\noalign{\smallskip}
Date & $F_{H\gamma}$ & $EW_{H\gamma}$ & $F_{H\delta}$ & $EW_{H\delta}$ \\ 
\noalign{\smallskip}
yyyy-mm-dd &  $\times10^{-12}$ [erg s$^{-1}$ cm$^{-2}$] & [\AA] & $\times10^{-12}$ [erg s$^{-1}$ cm$^{-2}$] & [\AA] \\
\noalign{\smallskip}\hline
\noalign{\smallskip}
2011-06-08 & $1.89 \pm 0.05$ & $-23.21 \pm 0.04$ & $1.31 \pm 0.02$ & $-16.70 \pm 0.02$ \\
2011-06-11 & $2.29 \pm 0.04$ & $-23.31 \pm 0.04$ & $1.52 \pm 0.01$ & $-14.91 \pm 0.02$ \\
2011-06-12 & $3.18 \pm 0.05$ & $-23.26 \pm 0.04$ & $2.11 \pm 0.02$ & $-12.99 \pm 0.02$ \\
2011-06-14 & $1.82 \pm 0.05$ & $-22.77 \pm 0.04$ & $1.29 \pm 0.02$ & $-17.25 \pm 0.02$ \\
2011-06-15 & $2.00 \pm 0.03$ & $-21.38 \pm 0.04$ & $1.36 \pm 0.01$ & $-13.64 \pm 0.03$ \\
2011-06-16 & $1.85 \pm 0.05$ & $-18.08 \pm 0.04$ & $1.28 \pm 0.02$ & $-12.27 \pm 0.02$ \\
2011-06-17 & $2.13 \pm 0.05$ & $-20.23 \pm 0.04$ & $1.41 \pm 0.02$ & $-12.25 \pm 0.02$ \\
2011-06-18 & $2.57 \pm 0.04$ & $-30.81 \pm 0.04$ & $1.67 \pm 0.02$ & $-17.89 \pm 0.02$ \\
2011-06-20 & $1.65 \pm 0.06$ & $-17.77 \pm 0.04$ & $1.18 \pm 0.02$ & $-13.47 \pm 0.02$ \\
2011-06-21 & $1.47 \pm 0.05$ & $-18.60 \pm 0.04$ & $1.04 \pm 0.01$ & $-14.13 \pm 0.02$ \\
2011-06-22 & $2.21 \pm 0.04$ & $-22.73 \pm 0.04$ & $1.50 \pm 0.01$ & $-14.58 \pm 0.02$ \\
2011-06-23 & $2.28 \pm 0.04$ & $-21.88 \pm 0.04$ & $1.49 \pm 0.01$ & $-12.89 \pm 0.02$ \\
\noalign{\smallskip}
\hline
\end{tabular}
\caption{Integrated fluxes and equivalent widths of the \ion{H}{i} lines and their associated errors. }
\label{table:H_fluxes}
\end{table}

\begin{table}[h]
\centering
\begin{tabular}{l c c c c c c c c}
\hline\hline  
\noalign{\smallskip}
Date & $F_{\ion{He}{i}\,\,5875}$ & $EW_{\ion{He}{i}\,\,5875}$ & $F_{\ion{He}{i}\,\,6678}$ & $EW_{\ion{He}{i}\,\,6678}$ \\ 
\noalign{\smallskip}
yyyy-mm-dd & $\times10^{-13}$ [erg s$^{-1}$ cm$^{-2}$] & [\AA] & $\times10^{-13}$ [erg s$^{-1}$ cm$^{-2}$] & [\AA] \\
\noalign{\smallskip}\hline
\noalign{\smallskip}

2011-06-08 & $4.41 \pm 0.01$ & $-4.05 \pm 0.01$ & $2.18 \pm 0.06$  & $-1.65 \pm 0.02$ \\ 
2011-06-11 & $6.10 \pm 0.04$ & $-4.76 \pm 0.02$ & $2.97 \pm 0.05$  & $-1.91 \pm 0.02$ \\ 
2011-06-12 & $8.98 \pm 0.06$ & $-5.05 \pm 0.02$ & $4.91 \pm 0.06$  & $-2.38 \pm 0.02$ \\
2011-06-14 & $5.10 \pm 0.03$ & $-4.45 \pm 0.02$ & $2.56 \pm 0.06$  & $-1.84 \pm 0.03$ \\
2011-06-15 & $5.23 \pm 0.03$ & $-4.06 \pm 0.02$ & $2.71 \pm 0.05$  & $-1.76 \pm 0.02$ \\
2011-06-16 & $5.29 \pm 0.04$ & $-4.01 \pm 0.02$ & $2.80 \pm 0.06$  & $-1.79 \pm 0.02$ \\
2011-06-17 & $5.86 \pm 0.04$ & $-4.14 \pm 0.02$ & $3.04 \pm 0.05$  & $-1.79 \pm 0.02$ \\
2011-06-18 & $6.38 \pm 0.04$ & $-5.06 \pm 0.02$ & $3.15 \pm 0.06$  & $-1.99 \pm 0.02$ \\
2011-06-20 & $4.46 \pm 0.01$ & $-3.86 \pm 0.01$ & $1.89 \pm 0.05$  & $-1.36 \pm 0.02$ \\
2011-06-21 & $3.75 \pm 0.01$ & $-3.43 \pm 0.02 $ & $1.68 \pm 0.04$  & $-1.25 \pm 0.02$ \\ 
2011-06-22 & $5.93 \pm 0.03$ & $-4.46 \pm 0.02$ & $2.99 \pm 0.05$  & $-1.88 \pm 0.02$ \\
2011-06-23 & $5.34 \pm 0.04$ & $-3.78 \pm 0.02$ & $2.60 \pm 0.04$  & $-1.56 \pm 0.02$ \\
\noalign{\smallskip}
\hline
\end{tabular}
\caption{Integrated fluxes and equivalent widths of \ion{He}{i}\,5875\,\AA\,and \ion{He}{i}\,6678\,\AA\,and their associated errors.}
\label{table:HeI_fluxes}
\end{table}

\section{Accretion-tracing line profiles}

Below are the line profiles of the \ion{H}{i} and \ion{He}{i} emission lines. Each night of observations is a different colour to better visualise the variability.

\begin{figure}
    \centering
    \includegraphics[width=1.0\linewidth]{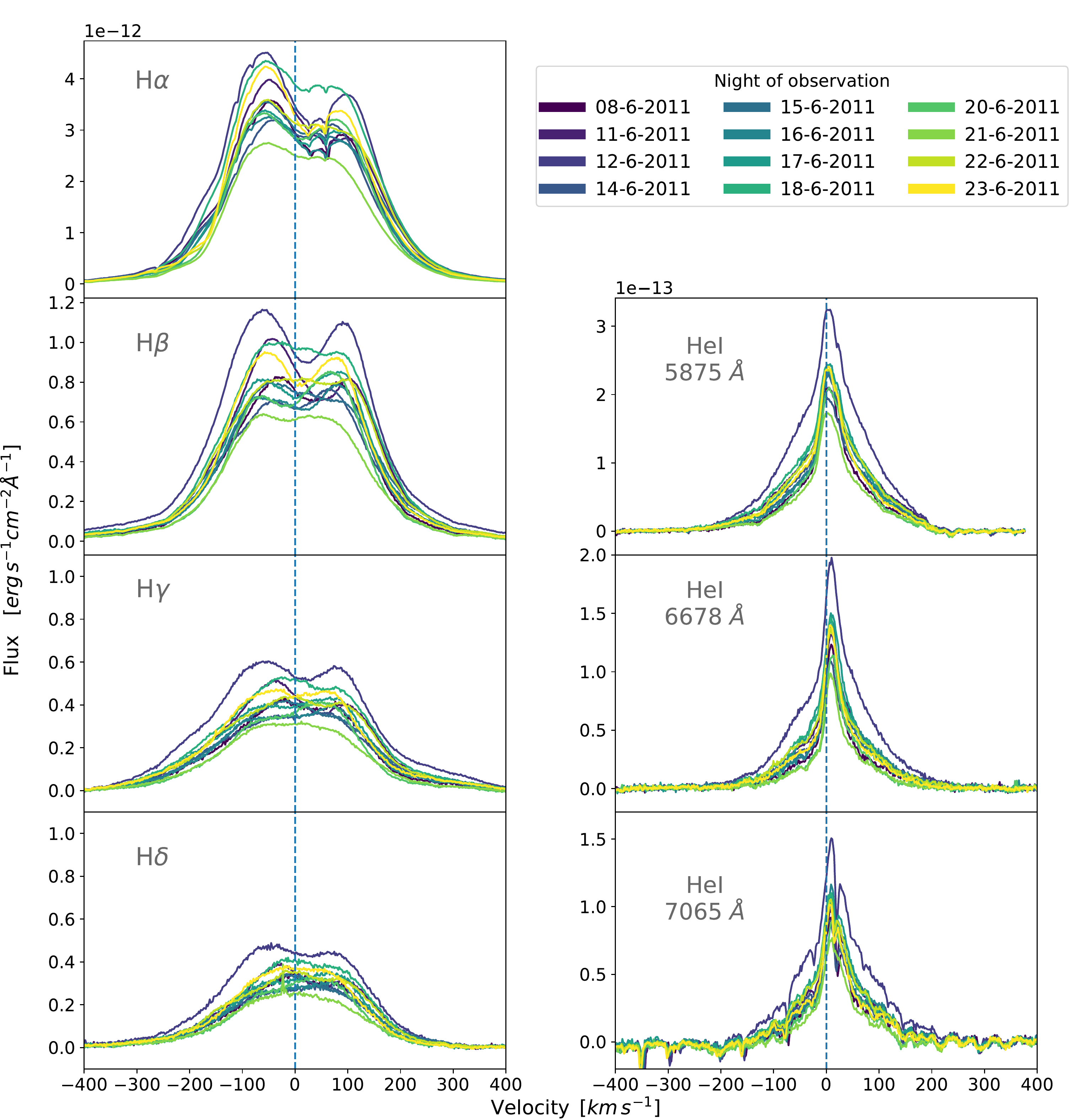}
    \caption{Accretion-tracing lines of \ion{H}{i} and \ion{He}{i} over the 12 epochs of ESPaDOnS observations spanning from 08 June 2011 to 23 June 2011. These spectra are continuum-subtracted and flux calibrated. They illustrate a night-to-night variability in the line emission from RU\,Lup.}
    \label{fig:acc_lines}
\end{figure}

\section{Method of calculating veiling}
\label{appendix:calc_veiling}

The excess continuum that magnetospheric accretion adds to a CTTSs optical spectrum has been investigated since \citet{Joy1949} first introduced the concept of veiling. With high resolution spectroscopy, it is possible to observe the photospheric absorption lines of the star and this excess due to accretion that effectively veils the spectrum. In \citet{Hartigan1989}, a quantitative method was developed to measure the amount of veiling present in a given CTTS. By comparing the spectrum of the accreting CTTS, with a spectrum that represents only the photosphere (no accretion present), the following equation is produced:
\begin{equation}
    O(\lambda_i) = A\,[S(\lambda_i) + V(\lambda_i)],
\end{equation}
where $O(\lambda_i)$ is the observed spectrum of the CTTS, $S(\lambda_i)$ is the template or synthetic spectrum containing only the photospheric flux, $V(\lambda_i)$ is the veiling spectrum (the excess due to accretion) and $A$ is a normalisation constant. As it is difficult to measure the veiling across the entire spectrum, it can be measured within a specific wavelength range:
\begin{equation}
    O(\lambda_i) = A_j\,[S(\lambda_i) + k_j],
\end{equation}
where $k_j$ is the excess flux in a specific wavelength interval $j$ (in \citet{Hartigan1989} intervals of $10-15$\,\AA\,are used) and $A_j$ is the normalisation at this interval.

The dimensionless quantity veiling commonly used is represented as $r_j$:
\begin{equation}
    r_j = \frac{k_j}{S_j}.
\end{equation}
\citet{Hartigan1989} defines $S_{cont}$ as the estimated continuum level of the template star used for comparison at the wavelength range $j$:

\begin{equation}
    r_j = \frac{F_\text{excess}}{F_{\text{photosphere}}}.
\end{equation}
Therefore, the total flux observed in the CTTS is defined as, $F_{\text{total}} = F_{\text{photosphere}} + F_{\text{excess}}$. By measuring the amount of veiling in a CTTS, it is possible to quantify the amount of accretion activity happening in the star.

\section{Line profiles \& line decomposition}

Below are the line profile decompositions for the \ion{He}{i} emission lines.

\begin{figure}[h]
    \centering
    \includegraphics[width=0.49\linewidth]{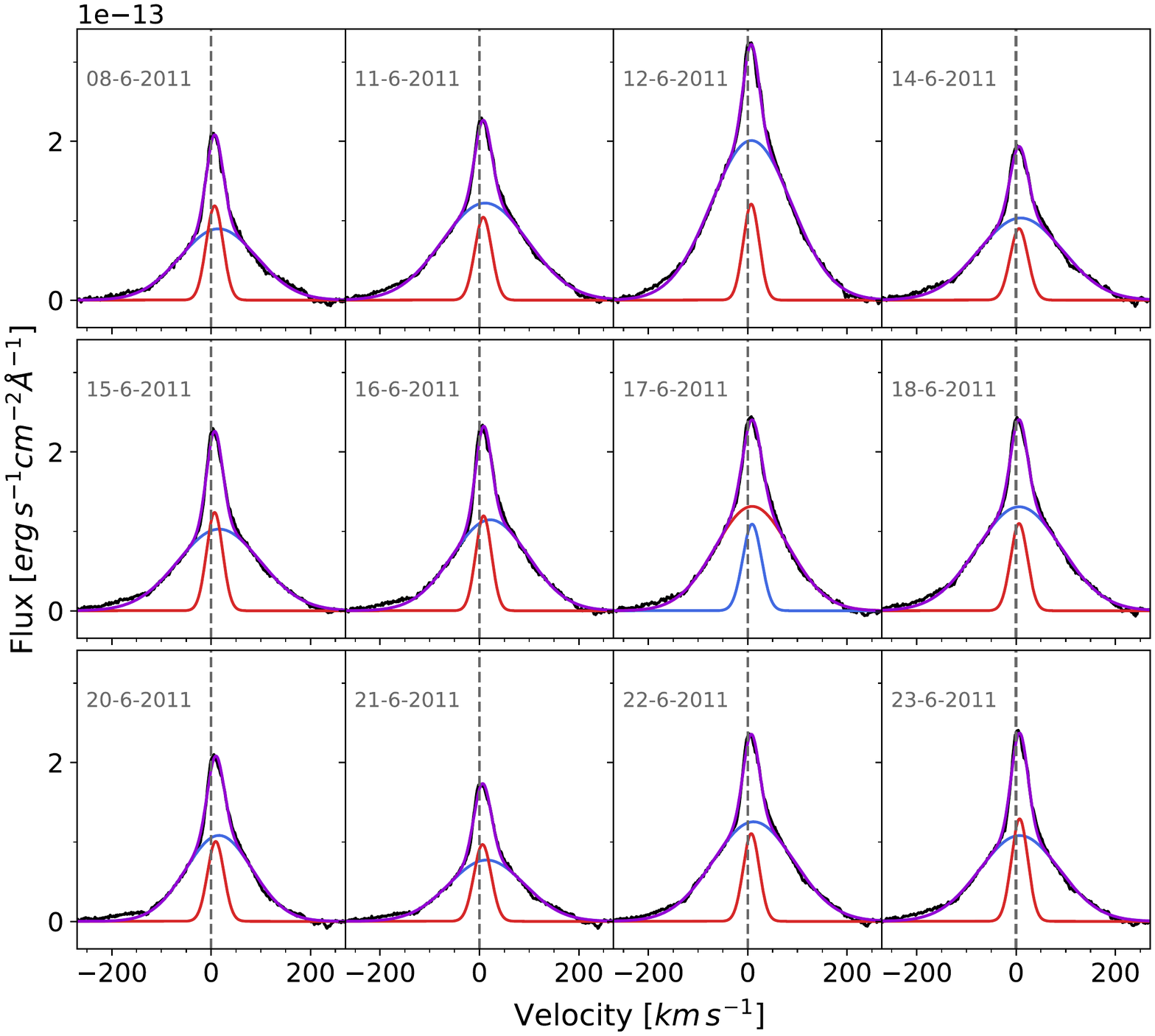}
    \includegraphics[width=0.49\linewidth]{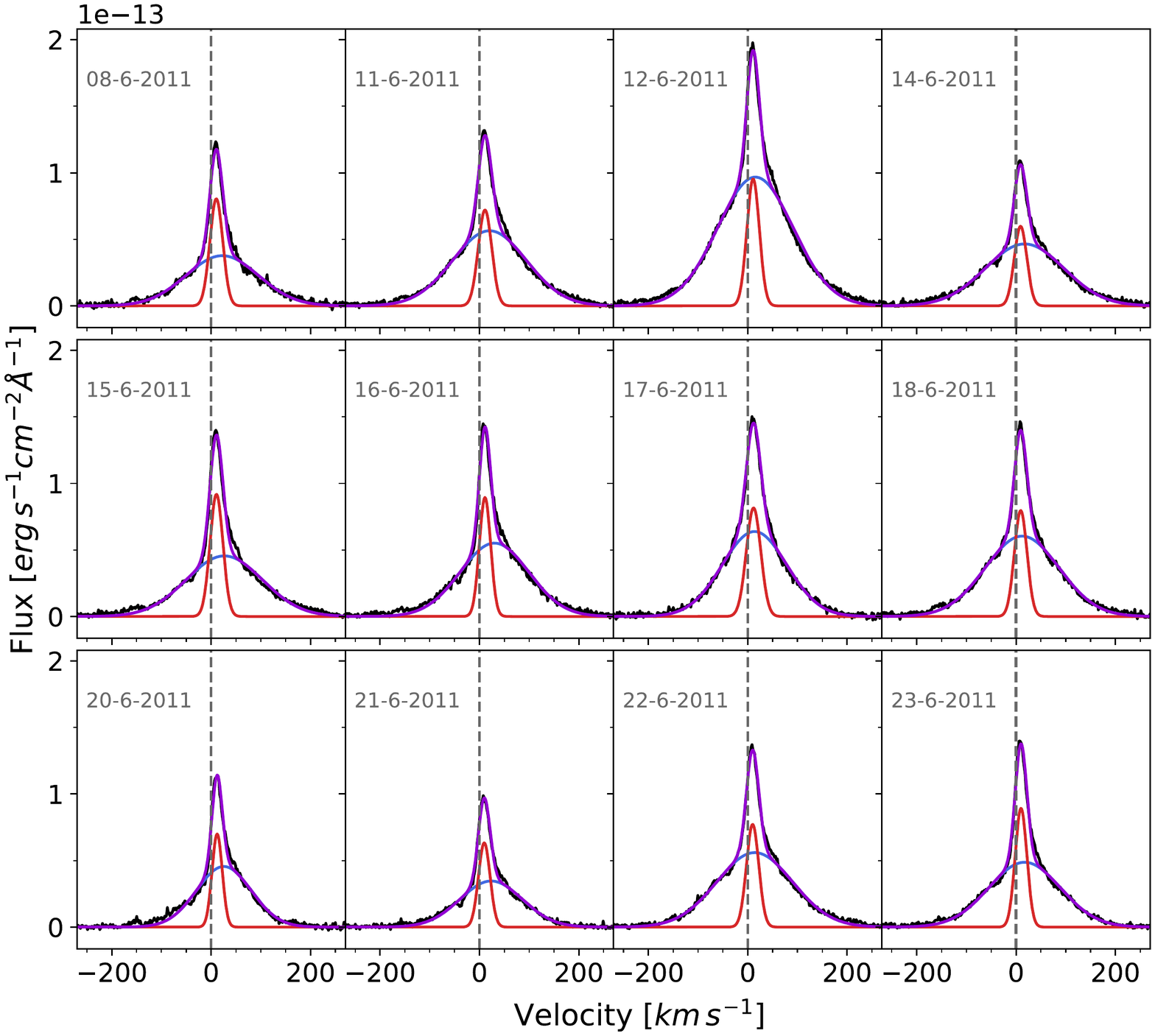}
    \caption{Gaussian decompositions of the \ion{He}{i}\,5875\,\AA\,(left-hand plots) and 6678\,\AA\,(right-hand plots) emission lines for each night of observations. The spectra are continuum-subtracted and flux calibrated.}
\label{fig:heI_decomp}
\end{figure}
\begin{figure}[h]
    \centering
    \includegraphics[width=0.45\linewidth]{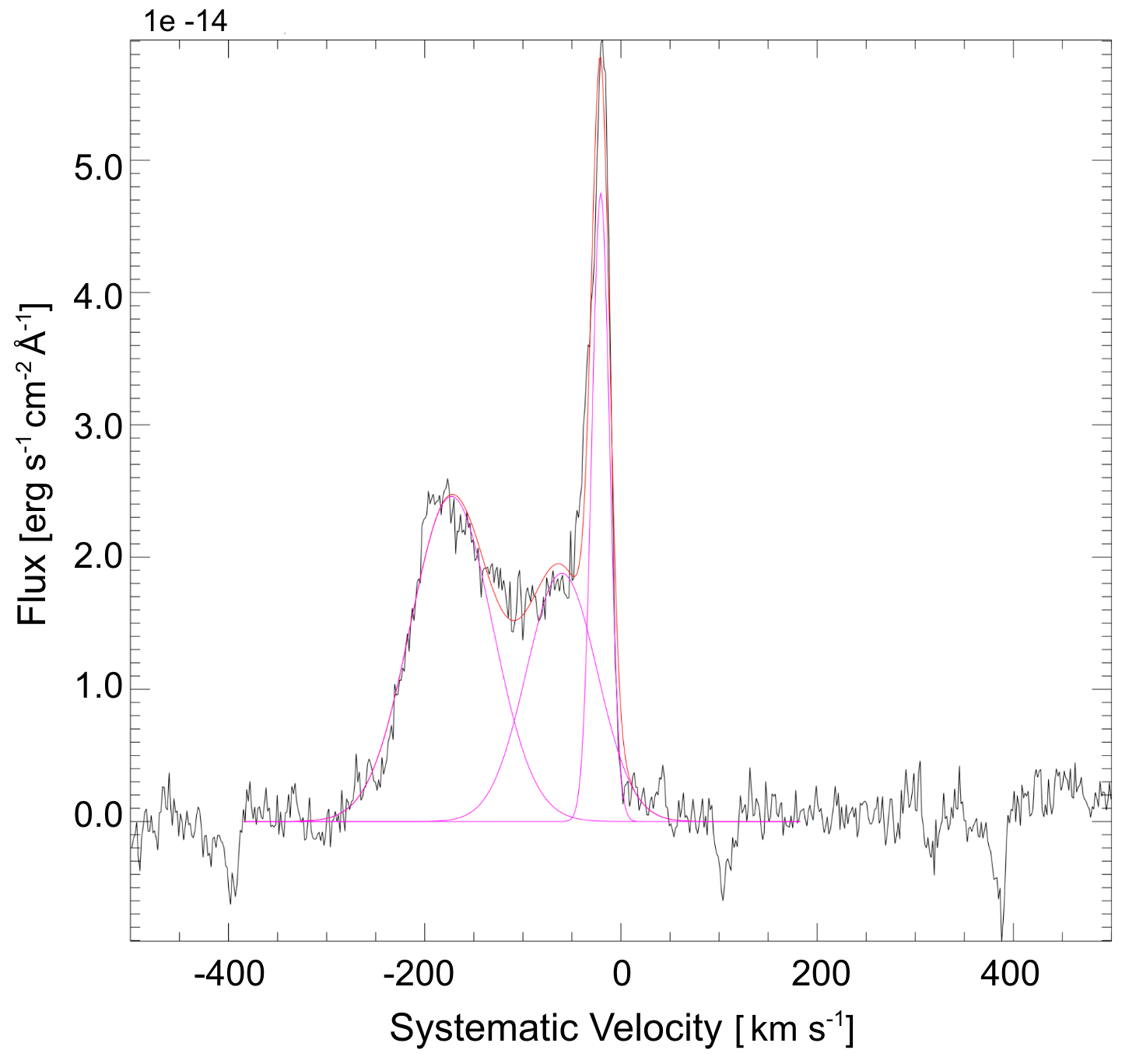}
    \caption{Example of one of the Gaussian decompositions of the [\ion{S}{ii}]\,6730\,\AA\, lines from 14 June 2011. The spectra are continuum-subtracted and flux calibrated.}
    \label{fig:outflow_line_fits}
\end{figure}

\newpage
\section{Wavelength-dependence of $L_{\text{\,acc}}$ estimates}
\begin{figure}[h]
    \centering
    \includegraphics[width=0.75\linewidth]{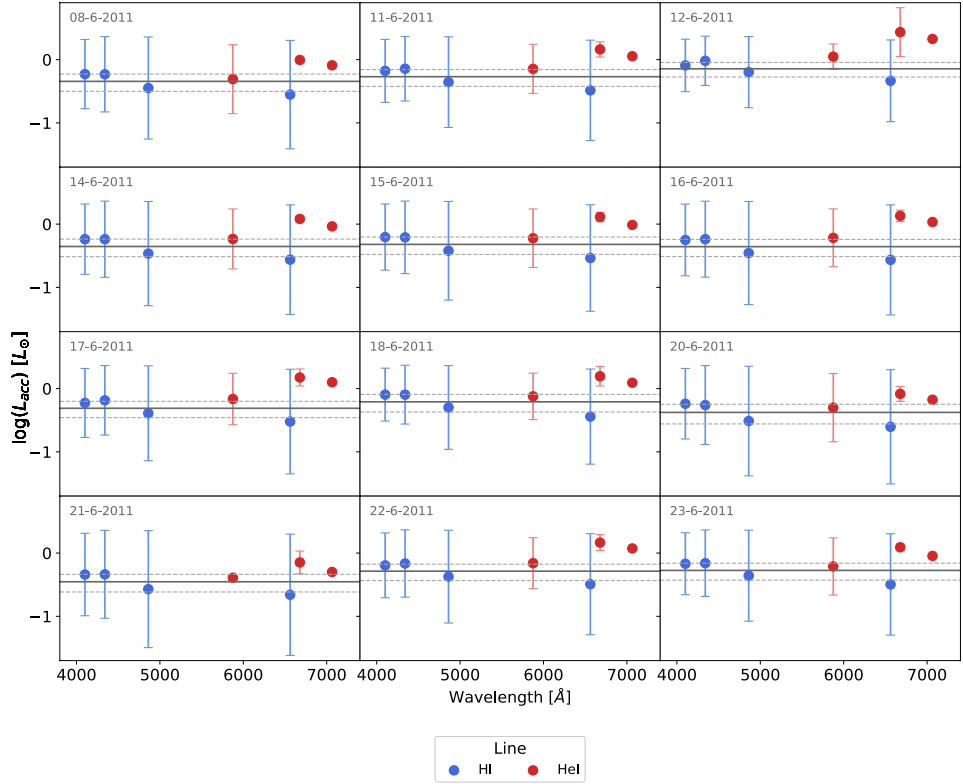}
    \caption{Plots searching for any possible wavelength dependence of the estimate of $L_{\text{\,acc}}$ for each of the 12 epochs. None is found. The horizontal line represents the $L_{\text{\,acc}}$ used as a result, which has been derived from an average of only the \ion{H}{i}'s values for $L_{\text{\,acc}}$ (blue dots). The \ion{He}{i} (red dots) lines are also shown.}
    \label{fig:man_plot}
\end{figure}

\end{appendix}

\end{document}